\documentclass[reprint,superscriptaddress,amsmath,longbibliography,amssymb,prl]{revtex4-2}

\usepackage{graphicx}
\usepackage{dcolumn}
\usepackage{bm}
\usepackage{multirow}
\usepackage{array}
\usepackage[usenames,dvipsnames]{xcolor}
\definecolor{nblue}{rgb}{0.0, 0.0, 1.0}
\definecolor{magenta}{rgb}{0.79, 0.08, 0.48}
\usepackage[colorlinks,linkcolor=nblue,urlcolor=magenta,citecolor=magenta,plainpages=false,pdfpagelabels,breaklinks]{hyperref}

\newcommand{\beq}{\begin{equation}}
\newcommand{\eeq}{\end{equation}}
\newcommand{\bea}{\begin{eqnarray}}
\newcommand{\eea}{\end{eqnarray}}
\newcommand{\non}{\nonumber}
\newcommand{\bk} { \bm{k} }
\newcommand{\bq} { \bm{q} }
\newcommand{\bd} { \bm{d} }
\newcommand{\eqn}[1] {Eq.~(\ref{#1})}
\newcommand{\fig}[1]{Fig.~\ref{#1}}
\newcommand{\mylabel}[1]{\label{#1}}
\newcommand{\tab}[1]{Table~\ref{#1}}

\newcommand{\titlename}{Time-Reversal Symmetry Breaking Superconductivity in Three-Dimensional Dirac Semimetallic Silicides}

\begin{document}


\title{\titlename}


\author{Sudeep\ K.\ Ghosh}\email[Electronic address: ]{S.Ghosh@kent.ac.uk}
\affiliation{School of Physical Sciences, University of Kent, Canterbury CT2 7NH, United Kingdom.}

\author{P.\ K.\ Biswas}\thanks{These authors contributed equally: S. K. Ghosh and P. K. Biswas.}
\affiliation{ISIS Pulsed Neutron and Muon Source, STFC Rutherford Appleton Laboratory, Harwell Campus, Didcot, Oxfordshire OX11 0QX, United Kingdom.}

\author{Chunqiang Xu}
\affiliation{Key Laboratory of Quantum Precision Measurement of Zhejiang Province, Department of Applied Physics, Zhejiang University of Technology, Hangzhou 310023, China}
\affiliation{School of Physics and Key Laboratory of MEMS of the Ministry of Education, Southeast University, Nanjing 211189, China}

\author{B. Li}
\affiliation{Information Physics Research Center, Nanjing University of Posts and Telecommunications, Nanjing 210023, China}

\author{J. Z. Zhao}
\affiliation{Co-Innovation Center for New Energetic Materials, Southwest University of Science and Technology, Mianyang, China}

\author{A.\ D.\ Hillier}\email[Electronic address: ]{adrian.hillier@stfc.ac.uk}
\affiliation{ISIS Pulsed Neutron and Muon Source, STFC Rutherford Appleton Laboratory, Harwell Campus, Didcot, Oxfordshire OX11 0QX, United Kingdom.}

\author{Xiaofeng Xu}\email[Electronic address: ]{xuxiaofeng@zjut.edu.cn}
\affiliation{Key Laboratory of Quantum Precision Measurement of Zhejiang Province, Department of Applied Physics, Zhejiang University of Technology, Hangzhou 310023, China}


\date{\today}

\begin{abstract}
Superconductors with broken time-reversal symmetry represent arguably one of the most promising venues for realizing highly sought-after topological superconductivity that is vital to fault-tolerant quantum computation. 
Here, by using extensive muon-spin relaxation and rotation measurements, we report that the isostructural silicide superconductors (Ta, Nb)OsSi spontaneously break time-reversal symmetry at the superconducting transition while surprisingly showing a fully-gapped superconductivity characteristic of conventional superconductors. The first-principles calculations show that (Ta, Nb)OsSi are three-dimensional Dirac semimetals protected by nonsymmorphic symmetries. Taking advantage of the exceptional low symmetry crystal structure of these materials, we have performed detailed theoretical calculations to establish that the superconducting ground state for both (Ta, Nb)OsSi is most likely a nonunitary triplet state. 

\end{abstract}

\maketitle

\section{Introduction}

Dirac or Weyl semimetals have attracted significant research interest due to their exceptional physical properties arising from the topologically protected gapless electronic excitations~\cite{Armitage2018,Lv2021}. Three-dimensional (3D) Dirac semimetals are particularly interesting because they can induce novel topological phases when symmetries are broken, e.g. a Dirac semimetal transforms into a Weyl semimetal with time-reversal symmetry (TRS) breaking~\cite{Armitage2018}. However, research on topological semimetals to date has been primarily focused on characterizing the underlying nontrivial band topology, while their interplay with correlated electronic states, such as novel magnetism and unconventional superconductivity, remained largely an uncharted territory.  On the other hand,
an interesting class of unconventional superconductors includes the ones that spontaneously break TRS in the superconducting state~\cite{Ghosh2020a} yet otherwise have properties similar to conventional BCS-type superconductors. 
As a result, superconducting 3D Dirac semimetals that break TRS in the superconducting state represent a unique class of materials to realize novel topological superconductivity~\cite{sato2017} but are extremely rare.

A superconducting order parameter which breaks TRS is, generically, required to have multiple components with non-trivial phases in between~\cite{Ghosh2020a}. Such a multi-component order parameter arises from a multidimensional irreducible representation~\cite{Annett1990,sigrist1991} of the crystal point group of the material. However, it is usually difficult to unambiguously establish the structure of the superconducting order parameters for the TRS-breaking superconductors mainly due to two reasons: a) lack of sufficient knowledge of the electron pairing mechanism and b) highly symmetric crystal structures leading to many possibilities with similar low-temperature properties. This limits our ability to work by the process of elimination. For example, the point group $D_{4h}$ of Sr$_2$RuO$_4$ allows for 20 possibilities with weak spin-orbit coupling (SOC) and 2 possibilities with strong SOC, of TRS-breaking superconducting instabilities~\cite{Annett1990,sigrist1991}. In this regard, the superconductors LaNiC$_2$~\cite{Hillier2009,chen2013}, LaNiGa$_2$~\cite{Hillier2012,Weng2016,Ghosh2020b,Badger2021} and UTe$_2$~\cite{ran2019}  are exceptions due to their very low-symmetry crystal structure that leads only to a few symmetry-allowed superconducting order parameters. In contrast to LaNiC$_2$~\cite{chen2013} and LaNiGa$_2$~\cite{Weng2016,Ghosh2020b}, which show two full gaps, UTe$_2$ shows nodal behaviour~\cite{Tristin2019} in their respective TRS-breaking superconducting state.

The recently discovered Osmium-based silicide superconductors (Ta,Nb)OsSi~\cite{Benndorf2017} have a very low-symmetry crystal structure as well and weak electron-phonon coupling~\cite{Haque2018}. In this article, by a combination of multiple experimental techniques including muon-spin rotation and relaxation ($\mu$SR) and thermodynamic measurements along with a detailed theoretical analysis, we demonstrate that (Ta,Nb)OsSi belong to nonsymmorphic symmetry-protected 3D Dirac semimetals that spontaneously break TRS at the superconducting transition but behave as conventional superconductors otherwise. By means of symmetry analysis and model calculations, our observations are found to be consistent with a nonunitary triplet superconducting ground state, the verification of which shall stimulate further study, both experimentally and theoretically.

\section{Results and Discussion}

(Ta, Nb)OsSi crystallize in a TiNiSi-type centrosymmetric orthorhombic crystal structure and have similar physical and chemical properties~\cite{Benndorf2017}. We prepared polycrystalline samples of (Ta, Nb)OsSi using conventional solid state reaction method and systematically investigated their physical properties using detailed $\mu$SR measurements in zero-field (ZF), longitudinal-field (LF) and transverse-field (TF) modes; magnetic-susceptibility, specific-heat and electrical-resistivity measurements~\cite{SM}. The $\mu$SR measurements were performed using the MUSR spectrometer at the ISIS Pulsed Neutron and Muon Source, UK. The temperature dependence of the magnetic susceptibility, collected in zero-field-cooled mode on the same samples, is shown by the solid blue lines on the right axis of \fig{fig:ZF}\textbf{b} (\fig{fig:ZF}\textbf{d}) for TaOsSi (NbOsSi). It indicates bulk superconductivity with $T_{\rm c} \approx 5.5$ K in TaOsSi and $T_{\rm c} \approx 3.1$ K in NbOsSi.\\

\begin{figure}[ht]
\begin{center}
\includegraphics[width=\columnwidth]{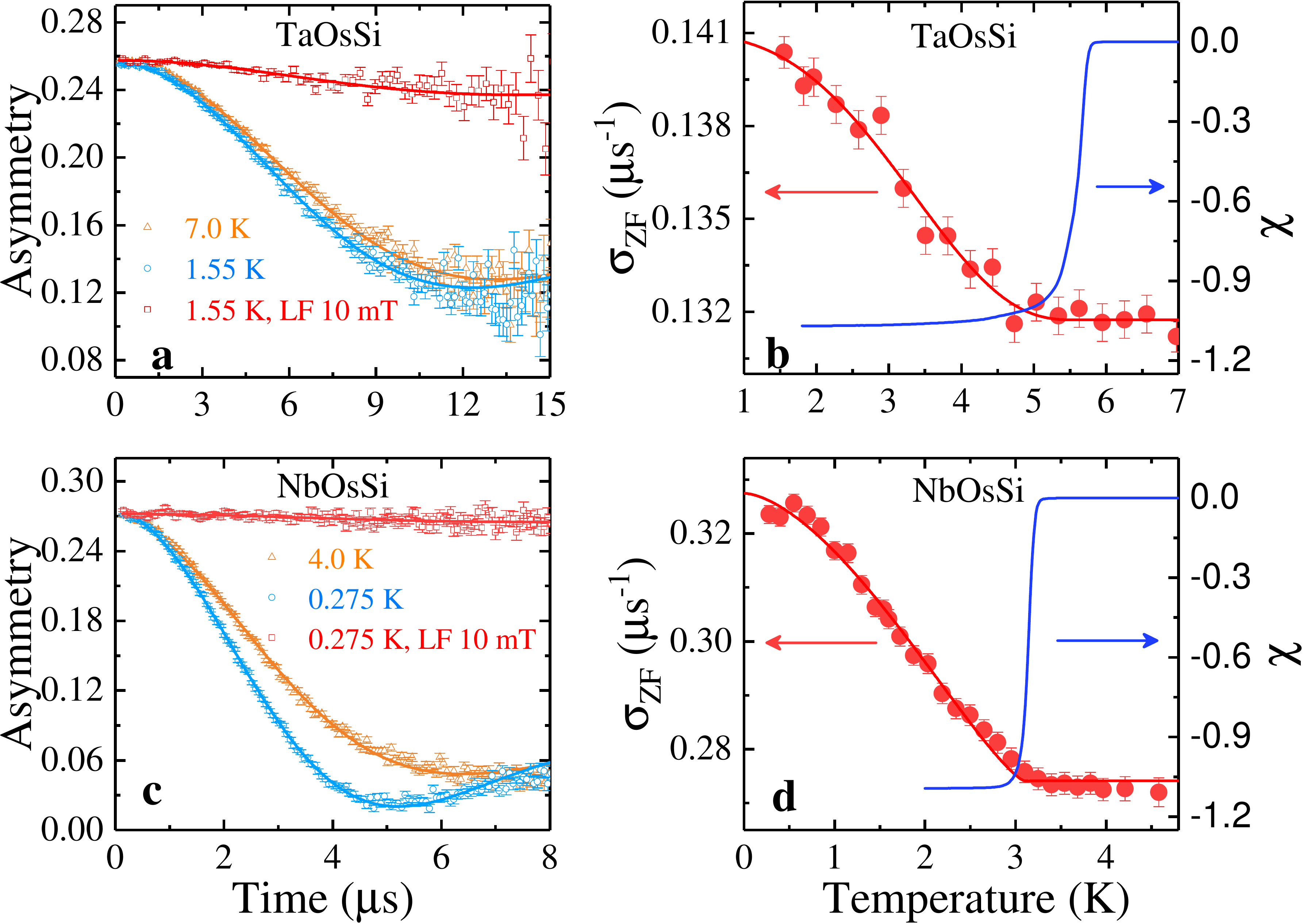}
\caption{\textbf{Time-reversal symmetry breaking observed in (Ta, Nb)OsSi by ZF-$\mu$SR measurements.} \textbf{a} and \textbf{c}: ZF-$\mu$SR time spectra collected above and below $T_{\rm c}$ for TaOsSi and NbOsSi respectively. The solid lines are the fits to the data using Eq.~\ref{eq:KT_ZFequation}. \textbf{b} and \textbf{d}: The temperature dependence of the muon spin relaxation rate $\sigma_{\rm ZF}$ of TaOsSi and NbOsSi respectively. Both show a clear systematic increase in $\sigma_{\rm ZF}$ just below $T_{\rm c}$. The corresponding $T_{\rm c}$-s are shown in the diamagnetic shifts in the magnetic susceptibility data of the two materials by the solid blue lines. The error bars shown are the standard deviations in the respective measurements.}
\mylabel{fig:ZF}
\end{center}
\end{figure}

\noindent\textbf{ZF-$\mu$SR:} ZF-$\mu$SR measurements were performed in search for spontaneous magnetic fields that can appear in the superconducting state leading to breaking of TRS. \fig{fig:ZF}\textbf{a} and \fig{fig:ZF}\textbf{c} show the ZF-$\mu$SR time spectra for TaOsSi and NbOsSi respectively, collected above and below the respective $T_{\rm c}$-s. A clear increase in the muon-spin relaxation rate in the superconducting state compared to the normal state is evident from both figures. The ZF-$\mu$SR time spectra over a range of temperatures across the $T_{\rm c}$ for both materials were collected. The data were fitted using a Gaussian Kubo-Toyabe relaxation function~\cite{Kubo1981} $\mathcal{G}(t) = \frac{1}{3}+\frac{2}{3}\left(1-\sigma_{\rm ZF}^2t^2\right){\exp}\left(-\sigma_{\rm ZF}^2t^2/2\right)$ multiplied by an exponential decay giving rise to the asymmetry function
\beq\mylabel{eq:KT_ZFequation}
A(t)= A(0) \mathcal{G}(t) {\exp}(-\lambda_{\rm ZF} t) + A_{\rm bg}.
\eeq
$A(0)$ and $A_{\rm bg}$ are the initial and background asymmetries of the ZF-$\mu$SR time spectra. $\sigma_{\rm ZF}$ and $\lambda_{\rm ZF}$ represent the muon spin relaxation rates originating from the presence of nuclear and electronic moments in the sample, respectively. In the fitting process, the electronic relaxation rate $\lambda_{\rm ZF}$ was found to be nearly temperature independent for both materials with small average values of $~ 0.0243(2)$ $\mu$s$^{-1}$ for TaOsSi and $~0.0633(5)$ $\mu$s$^{-1}$ for NbOsSi and hence was kept fixed. This indicates the absence of fast-fluctuating electronic moments. The nuclear relaxation rate $\sigma_{\rm ZF}(T)$ shown in \fig{fig:ZF}\textbf{b} (\fig{fig:ZF}\textbf{d}) for TaOsSi (NbOsSi), on the other hand, shows a clear systematic increase just below $T_{\rm c}$. The LF-$\mu$SR measurements performed under a field-cooled condition with a small field of 10 mT shown in Figs.~\ref{fig:ZF}\textbf{a} and \ref{fig:ZF}\textbf{c} clearly rule out the possibility of defect- or impurity-induced relaxations since the small field is enough to decouple the muon spins from the weak relaxation channels in both samples. This demonstrates that the increase in $\sigma_{\rm ZF}$ just below $T_{\rm c}$ is due to very weak fields which are static or quasi-static on the time-scale of muon life-time and are closely tied to the superconducting state, providing conclusive evidence of spontaneously broken TRS in the superconducting ground states of (Ta,Nb)OsSi. The spontaneous field estimated from the change $\Delta\sigma_{\rm ZF} = \sigma_{\rm ZF}(T \approx 0) - \sigma_{\rm ZF}(T>T_{\rm c})$ is $B_{\rm int} \approx \sqrt{2}\Delta\sigma_{\rm ZF}/\gamma_\mu=0.17$ G ($0.83$ G) for TaOsSi (NbOsSi) which is similar to other TRS breaking superconductors \cite{Ghosh2020a}. Here, $\gamma_{\mu}= 2\pi \times 135.5$~MHz/T is the muon gyromagnetic ratio~\cite{Sonier2000}.



\begin{figure*}[htb]
\begin{center}
\includegraphics[width=0.95\textwidth]{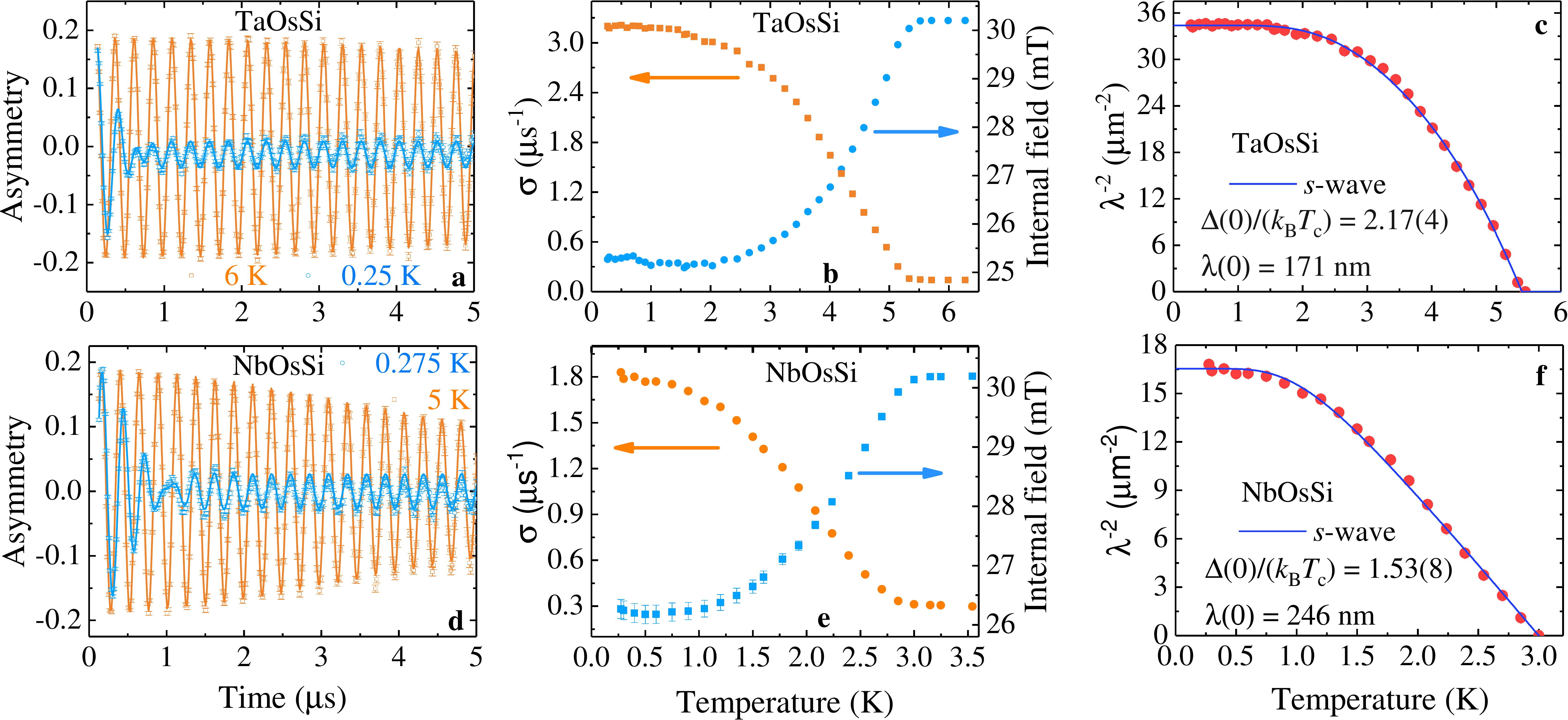}
\caption{\textbf{Characterizing the superconducting properties of (Ta, Nb)OsSi by TF-$\mu$SR measurements.} \textbf{a} and \textbf{d}: TF-$\mu$SR time spectra, collected above (red square) and below (blue circle) the $T_{\rm c}$ in a field-cooled transverse field of 30~mT for TaOsSi and NbOsSi respectively. The solid lines are the fits to the data using Eq.~\ref{Depolarization_Fit}. \textbf{b} and \textbf{e}: The temperature dependence of the extracted relaxation rate $\sigma$ (left axis) and internal field (right axis) for TaOsSi and NbOsSi respectively. \textbf{c} and \textbf{f}: The temperature dependence of the inverse magnetic penetration depth squared or equivalently the superfluid density $\rho_s \propto \lambda^{-2}(T)$ for TaOsSi and NbOsSi respectively. The solid lines are fits to the data with an isotropic single-gap $s$-wave model.}
 \label{fig:TF}
\end{center}
\end{figure*}

\noindent\textbf{TF-$\mu$SR:} To determine the superconducting gap symmetry of (Ta,Nb)OsSi, we have performed extensive TF-$\mu$SR measurements in a transverse field of 30~mT applied above $T_{\rm c}$ and cooled to the base temperature to stabilize a well-ordered flux-line-lattice in the mixed state of the superconductors. The TF-$\mu$SR asymmetry signals collected above and below $T_{\rm c}$ are shown in \fig{fig:TF}\textbf{a} (\fig{fig:TF}\textbf{d}) for TaOsSi (NbOsSi). For both materials, the asymmetry signals above $T_{\rm c}$ show very little relaxation due to the transverse component of weak nuclear moments present in these materials, while those below $T_{\rm c}$ show higher relaxation due to the added inhomogeneous field distribution of the flux-line-lattice.
The TF-$\mu$SR asymmetry signals were analysed using a Gaussian damped sinusoidal function plus a non-decaying oscillation that contributes to the muons stopping in the silver sample holder:
\begin{multline}
\label{Depolarization_Fit}
A_{TF}(t)=A(0)\exp\left(-\sigma^{2}t^{2}\right/2)\cos\left(\gamma_\mu \left\langle B\right\rangle t +\phi\right)\\
+A_{\rm bg}\cos\left(\gamma_\mu B_{\rm bg}t +\phi\right).
\end{multline}
Here $A(0)$ and $A_{\rm bg}$ are the initial sample and background asymmetries respectively, $\left\langle B\right\rangle$ and $B_{\rm bg}$ are the average internal and background magnetic fields respectively, $\phi$ is the shared phase offset and $\sigma$ is the depolarization rate of the muon spin precession signal originating from the variance of the magnetic-field distribution in the superconductor. \fig{fig:TF}\textbf{b} and \fig{fig:TF}\textbf{e} show the temperature dependence of the relaxation rate $\sigma$ (left axis) and internal field (right axis) of TaOsSi and NbOsSi respectively, extracted from the fits to the asymmetry signals using Eq.~\ref{Depolarization_Fit}. The internal fields at the muon sites show strong diamagnetic shifts below $T_{\rm c}$ for both materials, a clear indication of bulk superconductivity. The $\sigma=\left(\sigma^{2}_{\rm sc} + \sigma^{2}_{\rm nm}\right)^{\frac{1}{2}}$ includes contributions from both the flux-line-lattice $\sigma_{\rm sc}$ and a temperature-independent relaxation due to nuclear moments $\sigma_{\rm nm} = 0.146 \mu{s}^{-1}$ ($0.312\mu{s}^{-1}$) for TaOsSi (NbOsSi), determined from the average values of $\sigma$ collected above the respective $T_{\rm c}$-s where it is mostly temperature independent.

The London magnetic penetration depth $\lambda$ can be computed from $\sigma_{\rm sc}$ within a Ginzburg-Landau treatment of the vortex state in a superconductor in the limit of the applied field $H \ll H_{\rm c2}$ ~\cite{Brandt2003} as:
\begin{equation}
\frac{\sigma_{sc}\left(T\right)}{\gamma_\mu}=0.06091\frac{\Phi_0}{\lambda^{2}\left(T\right)},
\end{equation}
where $\Phi_0=2.068\times10^{-15}$~Wb is the flux quantum. The temperature dependence of $\lambda^{-2}$ extracted using the above equation for TaOsSi and NbOsSi are presented in \fig{fig:TF}\textbf{c} and \fig{fig:TF}\textbf{f} respectively. Since $\lambda^{-2}(T)$ is a measure of the superfluid density $\rho_s \propto \lambda^{-2} \propto n_{\rm s}/m^*$ ($n_{\rm s}$ is the charge carrier concentration, and $m^*$ is the effective mass of the charge carriers), it bears signatures of the symmetry of the superconducting gap. We note from \fig{fig:TF}\textbf{b} and \fig{fig:TF}\textbf{e} that the superfluid density of both materials shows saturation below ${T_{\rm c}}/3$ which indicates the absence of low-lying excited states close to zero temperature, a hallmark of node-less superconductivity.

\begin{figure*}[htb]
\begin{center}
\includegraphics[width=0.98\textwidth]{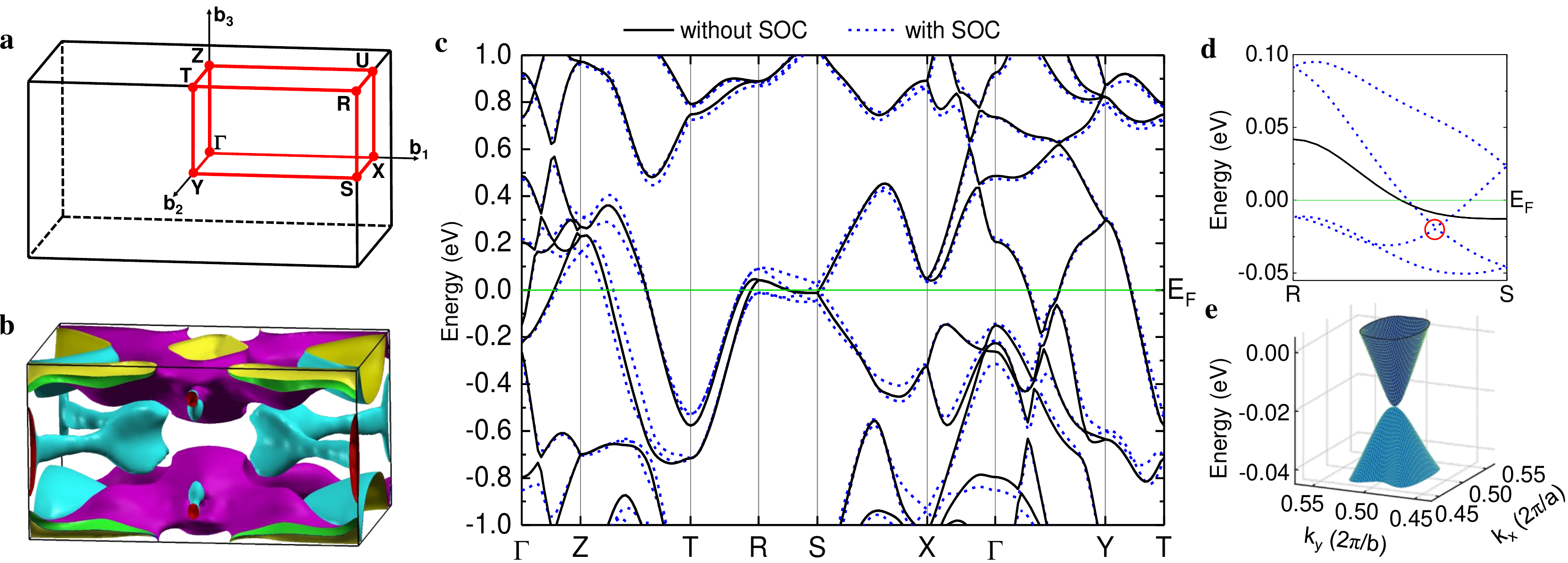}
\caption{\textbf{Band structure of NbOsSi.} \textbf{a}) First Brillouin zone with the high symmetry points marked. \textbf{b}) Combined view of all the four Fermi surface sheets of NbOsSi without SOC. The large parallel sections of the Fermi surface sheets are clearly visible. \textbf{c}) The band structure of NbOsSi with and without considering SOC. \textbf{d}) Enlarged view of the band structure along the RS-direction. The Dirac point is marked by the red circle. \textbf{e}) The dispersion close to the Dirac point marked in \textbf{d}.}
\label{fig:FS_fig}
\end{center}
\end{figure*}

To understand the superconducting pairing symmetry, we analyse the temperature dependence of the superfluid density within the local London approximation ($\lambda(0) \gg \xi$, $\xi$ is the coherence length)~\cite{Prozorov2006} by:
\begin{equation}
\frac{\lambda^{-2}\left(T\right)}{\lambda^{-2}\left(0\right)}=1+2 \bigg\langle \int^{\infty}_{\Delta_{\bk}(T)}\left(\frac{\partial f}{\partial E}\right)\frac{ EdE }{\sqrt{E^2-|\Delta_{\bk}\left(T\right)|^2}}\bigg\rangle_{\rm FS}.
\end{equation}
Here, $\Delta_{\bk}(T)$ is the form of the gap function for a given pairing model, $f(E,T)=\left[1+\exp\left(E/k_{\rm B}T\right)\right]^{-1}$ is the Fermi function and $\langle\rangle_{\rm FS}$ represents an average over a spherical Fermi surface. For an isotropic single gap $s$-wave model, $\Delta_{\bk}(T)$ is independent of $\bk$ and its temperature dependence is given by~\cite{Carrington2003}
\beq\label{eqn:temp_dep}
\Delta(T) = \Delta(0) \tanh\left[1.82\left\{1.018\left(T_{\rm c}/T-1\right)\right\}^{0.51}\right].
\eeq
The solid lines in the \fig{fig:TF}\textbf{c} and \fig{fig:TF}\textbf{f} show that the superfluid density can be fitted quite well with an isotropic single-gap $s$-wave model both for TaOsSi and NbOsSi respectively. We note that the values of the fitting parameter $\frac{\Delta(0)}{k_B T_{\rm c}}$ for both (Ta,Nb)OsSi are close to its weak-coupling BCS limit value.\\

\noindent\textbf{Band structure and specific heat:} The space group of (Ta,Nb)OsSi is Pnma (no. 62) and the point group is $D_{2h}$. The first Brillouin zone with the high-symmetry directions marked is shown in \fig{fig:FS_fig}\textbf{a}. Pnma is a nonsymmorphic space group having three glide planes: $G_1 = \{m_{(0,1,0)}|t_{(0,1/2,0)} \}$, $G_2 = \{m_{(0,0,1)}|t_{(1/2,0,1/2)} \}$ and $G_3 = \{m_{(1,0,0)}|t_{(1/2,1/2,1/2)} \}$ where $m$ and $t$ denote the mirror plane  and fractional translation parallel to the plane respectively. Two-fold degeneracies along the high symmetry lines $XS$, $XU$, $UR$ and $RS$ result from $G_2$ and that along $YS$ and $UZ$ result from $G_3$~\cite{SM}. The band structure of NbOsSi (which is similar to that of TaOsSi~\cite{Haque2018,Xu2019}) computed using density functional theory within the generalized gradient approximation is shown in \fig{fig:FS_fig}\textbf{c} with and without the effect of SOC. We note that SOC leads to small yet finite splitting of the bands near the Fermi level with a maximum splitting $\sim 100$ meV near the $R$ point (maximum splitting $\sim 140$ meV near the $S$ point for TaOsSi~\cite{Xu2019}).

(Ta, Nb)OsSi are inherently multi-band systems with the Nb 4\textit{d}-orbitals (Ta 5\textit{d}-orbitals) and the Os 5\textit{d}-orbitals contributing the most to the density of states (DOS) at the Fermi level. There are four Fermi surface sheets (without SOC) with two of them contributing $\sim 80 \%$ to the DOS at the Fermi level~\cite{SM}. A combined view of all the four Fermi surface sheets of NbOsSi without SOC is shown in \fig{fig:FS_fig}\textbf{b}. 

The Kramer's theorem guarantees that all the electronic bands of non-magnetic centrosymmetric materials (Ta,Nb)OsSi are at least two-fold degenerate even in the presence of SOC. We find that (Ta,Nb)OsSi have four bulk Dirac points within $\sim 10-20$ meV energy window below the Fermi level~\cite{SM}. Two of these Dirac points lie on the $RS\bar{R}$-line ($\bar{R} \equiv -R$) and  are protected by the nonsymmorphic symmetry $G_2$ which leads to the additional two-fold degeneracy. A zoomed-in view of the band structure of NbOsSi along the $RS$ direction is shown in \fig{fig:FS_fig}\textbf{d} and the dispersion close to the Dirac point along this line is shown in \fig{fig:FS_fig}\textbf{e}. The other two Dirac points are, however, not protected by symmetry and the surface Fermi arcs are unfortunately not clearly distinguishable~\cite{SM}. Thus (Ta,Nb)OsSi are nonsymmorphic symmetry-protected Dirac semimetals expected to have characteristic spectroscopic and transport properties~\cite{Lv2021}.

As a result of the exceptionally low symmetry crystal structure of (Ta, Nb)OsSi, in the strong SOC limit, there are no symmetry allowed TRS breaking superconducting order parameters. In the weak SOC limit, while the relevant point group $D_{2h} \otimes SO(3)$ ($SO(3)$ is the group of spin rotations in three dimensions) has four symmetry allowed TRS breaking superconducting instabilities, all of them have nodes~\cite{Hillier2012}. Thus clearly all the symmetry-allowed superconducting instabilities of (Ta, Nb)OsSi in the effective single band picture~\cite{nonsymmnote} are inconsistent with the experimental observations. 

\begin{figure}[t]
\begin{center}
\includegraphics[width=0.9\columnwidth]{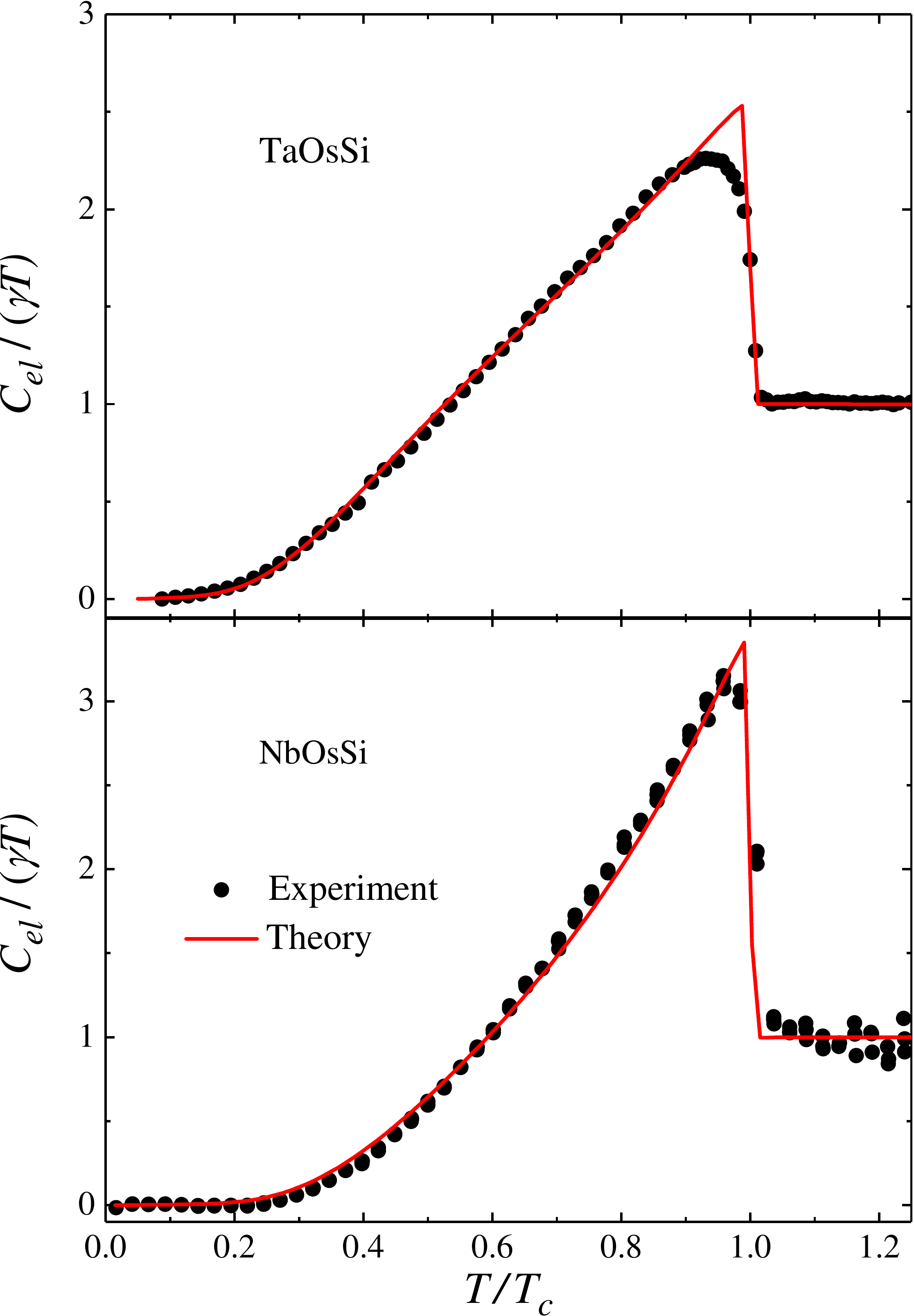}
\caption{\textbf{Electronic specific heat.} Temperature dependence of the experimentally measured electronic specific heat fitted with the theoretically computed specific heat in the INT state for the toy model with parameters $s/t = 0.1$, $\mu/t = -3.0$ and $\pmb{\eta} = \frac{1}{\sqrt{3}}(1, e^{i\pi/100}, e^{i 101\pi/100})$. The fitting parameters for TaOsSi case are $\Delta_0/(k_{\rm B} T_{\rm c}) = 2.20$ and $\alpha/t = 0.15$; and for NbOsSi case are $\Delta_0/(k_{\rm B} T_{\rm c}) = 2.63$ and $\alpha/t = 0.20$.}
\label{fig:Cv_fig}
\end{center}
\end{figure}

Motivated by the multi-band nature of (Ta, Nb)OsSi, we consider an internally antisymmetric nonunitary triplet (INT) superconducting state~\cite{Ghosh2020b} proposed in the case of LaNiGa$_2$, which has the same point group as (Ta, Nb)OsSi. Pairing in the INT state occurs between electrons on the same site but in two different orbitals in the nonunitary triplet channel and fermionic antisymmetry comes from orbital space. The pairing potential is $\hat{\Delta} = \hat{\Delta}_S \otimes \hat{\Delta}_B$ where the pairing potential in spin space is $\hat{\Delta}_S = (\bd.\pmb{\sigma})i \sigma_y$ and in orbital space is $\hat{\Delta}_B = i \tau_y$, with $\pmb{\sigma}$ ($\pmb{\tau}$) being the vector of Pauli matrices in spin space (orbital space). The triplet pairing is characterized by the $\bd$-vector: $\bd = \Delta_a \,\pmb{\eta}$ with $|\pmb{\eta}|^2 = 1$ which is nonunitary and is characterized by the real vector $\bq = i (\pmb{\eta} \times \pmb{\eta}^*) \neq 0$, and $\Delta_a$ is the pairing amplitude considered to be uniform to realize an isotropic gap observed in (Ta, Nb)OsSi. 

There are several extended regions inside the Brillouin zone of (Ta, Nb)OsSi where two of the Fermi surface sheets are parallel and close to each other as shown in \fig{fig:FS_fig}\textbf{b} for example. This feature is essential to stabilize the INT state and we model it by considering
a simple toy model with two bands $\epsilon_\pm (\bk) = \epsilon(\bk) \pm s$ emerging from two nearly degenerate effective orbitals rigidly shifted from each other by an energy $2s$. We consider a generic dispersion $\epsilon(\bk) = -2t [\cos(k_x) + \cos(k_y) + \cos(k_z)]$ with $t$ being a hopping energy scale and focus on the limit $s/t \ll 1$ implying small but finite splitting between the corresponding two Fermi surfaces. To take into account the effect of SOC present in these materials, we phenomenologically consider a Rashba-type SOC. Note that although (Ta, Nb)OsSi are globally centrosymmetric, inversion symmetry is broken locally at the (Ta, Nb) sites due to the TiNiSi-type structure~\cite{Araki2019} which can result in a Rashba-type SOC~\cite{Wu2017, Zhang2014}. Then the normal state Hamiltonian for the toy model is,
$\hat{\cal{H}}_0 = \sum_{\bk} \hat{c}^{\dagger}_{\bk} \cdot H_0(\bk) \cdot \hat{c}_{\bk}$,
defining $\hat{c}_{\bk} = \begin{bmatrix} \tilde{c}_{\uparrow, \bk} \\ \tilde{c}_{\downarrow, \bk} \end{bmatrix}$ with $\tilde{c}_{p,\bk} = \begin{bmatrix} c_{+, p , \bk} \\ c_{-, p, \bk} \end{bmatrix}$. $c_{\pm, p , \bk}$ is an electron annihilation operator in the $\pm$ band with spin $p = \uparrow$ and $\downarrow$, and
\beq\mylabel{eqn:normhammat}
H_0 (\bk) = \sigma_0 \otimes \begin{bmatrix} \xi_+(\bk) & 0 \\\\ 0 & \xi_{-}(\bk)\end{bmatrix} + (k_y \sigma_x - k_x \sigma_y) \otimes \alpha \tau_x
\eeq
where $\xi_{\pm}(\bk) = \epsilon_\pm(\bk) - \mu$ with $\mu$ being the chemical potential and $\sigma_0$ is the identity matrix in spin space. The second term in \eqn{eqn:normhammat} is a Rashba-type inter-orbital SOC of strength $\alpha$. Although, in general, both intra- and inter-orbital SOC terms should be present, we can fit the specific heat data well only in the limit of inter-orbital SOC strength much larger than the intra-orbital SOC strength, emphasizing the inter-orbital nature of the pairing in the INT state~\cite{SM}.

Using the Bogoliubov de-Gennes formalism~\cite{Ghosh2020a}, we computed the quasi-particle excitation spectrum for the toy model in the INT state considering the temperature dependence of $\Delta_a(T)$ in the form of \eqn{eqn:temp_dep}. The specific heat is then computed using the temperature dependent quasi-particle spectrum to fit the experimentally measured electronic specific heat after subtracting the phonon contribution~\cite{SM}. In the fitting shown in \fig{fig:Cv_fig}, we fixed $s/t \ll 1$ and $T_{\rm c}$ with the corresponding experimental values for (Ta,Nb)OsSi. Then there are only three fitting parameters: $\alpha/t$, $\pmb{\eta}$ and $\Delta_0/(k_{\rm B} T_{\rm c})$ with $\Delta_0 \equiv \Delta_a(0)$. \fig{fig:Cv_fig} shows that the electronic specific heat for both TaOsSi~\cite{oldspheat} and NbOsSi can be fitted very well with small SOC strengths ($\alpha/t \ll 1$) in the weak coupling limit. The corresponding INT ground state has $|\bq| = 0.03$ implying small but finite spin polarization which leads to the spontaneous magnetization in the superconducting state seen in the ZF-$\mu$SR experiments.\\

\section{Conclusions}

We have demonstrated through detailed $\mu$SR measurements that (Ta, Nb)OsSi belong to the rare class of TRS-breaking superconductors represented by LaNiC$_2$, LaNiGa$_2$ and UTe$_2$, all of which have very low-symmetry crystal structures providing a unique opportunity to constrain the superconducting order parameter from symmetries. While LaNiC$_2$~\cite{chen2013} and LaNiGa$_2$~\cite{Weng2016,Ghosh2020b} show two full gaps in the superconducting state arising from two spin channels, UTe$_2$~\cite{hayes2021,Tristin2019} shows point nodes and two different transitions in the superconducting state and only below the lowest transition temperature does the TRS breaking occur. In contrast, (Ta,Nb)OsSi show TRS breaking at $T_c$ but have a full gap and low temperature thermodynamic properties similar to a conventional BCS-type superconductor as evidenced from the TF-$\mu$SR and the specific heat data. Similarly, it will be interesting to investigate the other known isostructural superconductors in this family for possible TRS-breaking superconducting ground states, e.g. ZrOsSi which has contrasting properties than (Ta,Nb)OsSi and comparatively low  $T_c$~\cite{Zhong1986}, ZrIrSi and HfIrSi~\cite{Benndorf2017}.

By symmetry analysis and model calculations, the phenomenology of the superconducting properties of (Ta,Nb)OsSi are found to be overall consistent with a nonunitary triplet superconducting ground state. The presence of the Dirac points close to the Fermi level promotes interband pairing and further justifies the applicability of a minimal two-band toy model to describe the low energy normal state properties~\cite{Badger2021}. The nonsymmorphic symmetries present in (Ta,Nb)OsSi can allow for nodes in the order parameter at the Brillouin zone boundaries resulting in nodal superconducting states which are clearly incompatible with the full gap observed in the experiments. However, the nonsymmorphic symmetries can lead to degeneracies in the Bogoliubov quasi-particle bands which can result in topological superconductivity~\cite{Badger2021}. Further experimental studies in these materials, such as high-resolution ARPES, are highly desirable to confirm the presence of the symmetry-protected Dirac points close to the Fermi level, when the single crystals of these materials become available. (Ta,Nb)OsSi are therefore special symmetry-protected 3D Dirac semimetal superconductors that provide promising material platforms to investigate the rich physics arising from an interplay between topological Dirac fermions and unconventional TRS-breaking superconductivity. \\

\textit{\textbf{Acknowledgements}}: PKB gratefully acknowledges the ISIS Pulsed Neutron and Muon Source of the UK Science \& Technology Facilities Council (STFC) for access to the muon beam times. SKG acknowledges the Leverhulme Trust for support through the Leverhulme early career fellowship and thanks A. Agarwala, J. Quintanilla and T. Shiroka for discussions and comments on the manuscript. XX acknowledges the financial support from NSFC under Grant No. 11974061 and useful discussions with Xiangang Wan and Dong Qian.

\vspace{10pt}
\noindent\textbf{Author contributions:}
S.K.G., P.K.B and X.X conceived and initiated the project. C.X. and X.X. grew and characterized the samples used in this study. P.K.B and A.D.H. conducted the muon spin rotation and relaxation experiments. B.L. and J.Z.Z performed the first-principles calculations.  S.K.G constructed the theoretical understanding and wrote the paper with input from all the co-authors.


\bibliography{AOsSi}

\begin{thebibliography}{33}%
\makeatletter
\providecommand \@ifxundefined [1]{%
 \@ifx{#1\undefined}
}%
\providecommand \@ifnum [1]{%
 \ifnum #1\expandafter \@firstoftwo
 \else \expandafter \@secondoftwo
 \fi
}%
\providecommand \@ifx [1]{%
 \ifx #1\expandafter \@firstoftwo
 \else \expandafter \@secondoftwo
 \fi
}%
\providecommand \natexlab [1]{#1}%
\providecommand \enquote  [1]{``#1''}%
\providecommand \bibnamefont  [1]{#1}%
\providecommand \bibfnamefont [1]{#1}%
\providecommand \citenamefont [1]{#1}%
\providecommand \href@noop [0]{\@secondoftwo}%
\providecommand \href [0]{\begingroup \@sanitize@url \@href}%
\providecommand \@href[1]{\@@startlink{#1}\@@href}%
\providecommand \@@href[1]{\endgroup#1\@@endlink}%
\providecommand \@sanitize@url [0]{\catcode `\\12\catcode `\$12\catcode
  `\&12\catcode `\#12\catcode `\^12\catcode `\_12\catcode `\%12\relax}%
\providecommand \@@startlink[1]{}%
\providecommand \@@endlink[0]{}%
\providecommand \url  [0]{\begingroup\@sanitize@url \@url }%
\providecommand \@url [1]{\endgroup\@href {#1}{\urlprefix }}%
\providecommand \urlprefix  [0]{URL }%
\providecommand \Eprint [0]{\href }%
\providecommand \doibase [0]{https://doi.org/}%
\providecommand \selectlanguage [0]{\@gobble}%
\providecommand \bibinfo  [0]{\@secondoftwo}%
\providecommand \bibfield  [0]{\@secondoftwo}%
\providecommand \translation [1]{[#1]}%
\providecommand \BibitemOpen [0]{}%
\providecommand \bibitemStop [0]{}%
\providecommand \bibitemNoStop [0]{.\EOS\space}%
\providecommand \EOS [0]{\spacefactor3000\relax}%
\providecommand \BibitemShut  [1]{\csname bibitem#1\endcsname}%
\let\auto@bib@innerbib\@empty
\bibitem [{\citenamefont {Armitage}\ \emph {et~al.}(2018)\citenamefont
  {Armitage}, \citenamefont {Mele},\ and\ \citenamefont
  {Vishwanath}}]{Armitage2018}%
  \BibitemOpen
  \bibfield  {author} {\bibinfo {author} {\bibfnamefont {N.~P.}\ \bibnamefont
  {Armitage}}, \bibinfo {author} {\bibfnamefont {E.~J.}\ \bibnamefont {Mele}},\
  and\ \bibinfo {author} {\bibfnamefont {A.}~\bibnamefont {Vishwanath}},\
  }\bibfield  {title} {\bibinfo {title} {Weyl and \textsc{D}irac semimetals in
  three-dimensional solids},\ }\href
  {https://doi.org/10.1103/RevModPhys.90.015001} {\bibfield  {journal}
  {\bibinfo  {journal} {Rev. Mod. Phys.}\ }\textbf {\bibinfo {volume} {90}},\
  \bibinfo {pages} {015001} (\bibinfo {year} {2018})}\BibitemShut {NoStop}%
\bibitem [{\citenamefont {Lv}\ \emph {et~al.}(2021)\citenamefont {Lv},
  \citenamefont {Qian},\ and\ \citenamefont {Ding}}]{Lv2021}%
  \BibitemOpen
  \bibfield  {author} {\bibinfo {author} {\bibfnamefont {B.~Q.}\ \bibnamefont
  {Lv}}, \bibinfo {author} {\bibfnamefont {T.}~\bibnamefont {Qian}},\ and\
  \bibinfo {author} {\bibfnamefont {H.}~\bibnamefont {Ding}},\ }\bibfield
  {title} {\bibinfo {title} {Experimental perspective on three-dimensional
  topological semimetals},\ }\href
  {https://doi.org/10.1103/RevModPhys.93.025002} {\bibfield  {journal}
  {\bibinfo  {journal} {Rev. Mod. Phys.}\ }\textbf {\bibinfo {volume} {93}},\
  \bibinfo {pages} {025002} (\bibinfo {year} {2021})}\BibitemShut {NoStop}%
\bibitem [{\citenamefont {Ghosh}\ \emph
  {et~al.}(2020{\natexlab{a}})\citenamefont {Ghosh}, \citenamefont {Smidman},
  \citenamefont {Shang}, \citenamefont {Annett}, \citenamefont {Hillier},
  \citenamefont {Quintanilla},\ and\ \citenamefont {Yuan}}]{Ghosh2020a}%
  \BibitemOpen
  \bibfield  {author} {\bibinfo {author} {\bibfnamefont {S.~K.}\ \bibnamefont
  {Ghosh}}, \bibinfo {author} {\bibfnamefont {M.}~\bibnamefont {Smidman}},
  \bibinfo {author} {\bibfnamefont {T.}~\bibnamefont {Shang}}, \bibinfo
  {author} {\bibfnamefont {J.~F.}\ \bibnamefont {Annett}}, \bibinfo {author}
  {\bibfnamefont {A.~D.}\ \bibnamefont {Hillier}}, \bibinfo {author}
  {\bibfnamefont {J.}~\bibnamefont {Quintanilla}},\ and\ \bibinfo {author}
  {\bibfnamefont {H.}~\bibnamefont {Yuan}},\ }\bibfield  {title} {\bibinfo
  {title} {Recent progress on superconductors with time-reversal symmetry
  breaking},\ }\href {https://doi.org/10.1088/1361-648X/abaa06} {\bibfield
  {journal} {\bibinfo  {journal} {J. Phys. Condens. Matter}\ }\textbf {\bibinfo
  {volume} {33}},\ \bibinfo {pages} {033001} (\bibinfo {year}
  {2020}{\natexlab{a}})}\BibitemShut {NoStop}%
\bibitem [{\citenamefont {Sato}\ and\ \citenamefont {Ando}(2017)}]{sato2017}%
  \BibitemOpen
  \bibfield  {author} {\bibinfo {author} {\bibfnamefont {M.}~\bibnamefont
  {Sato}}\ and\ \bibinfo {author} {\bibfnamefont {Y.}~\bibnamefont {Ando}},\
  }\bibfield  {title} {\bibinfo {title} {Topological superconductors: a
  review},\ }\href {https://doi.org/10.1088/1361-6633/aa6ac7} {\bibfield
  {journal} {\bibinfo  {journal} {Rep. Prog. Phys.}\ }\textbf {\bibinfo
  {volume} {80}},\ \bibinfo {pages} {076501} (\bibinfo {year}
  {2017})}\BibitemShut {NoStop}%
\bibitem [{\citenamefont {Annett}(1990)}]{Annett1990}%
  \BibitemOpen
  \bibfield  {author} {\bibinfo {author} {\bibfnamefont {J.~F.}\ \bibnamefont
  {Annett}},\ }\bibfield  {title} {\bibinfo {title} {Symmetry of the order
  parameter for high-temperature superconductivity},\ }\href
  {https://doi.org/10.1080/00018739000101481} {\bibfield  {journal} {\bibinfo
  {journal} {Adv. Phys.}\ }\textbf {\bibinfo {volume} {39}},\ \bibinfo {pages}
  {83} (\bibinfo {year} {1990})}\BibitemShut {NoStop}%
\bibitem [{\citenamefont {Sigrist}\ and\ \citenamefont
  {Ueda}(1991)}]{sigrist1991}%
  \BibitemOpen
  \bibfield  {author} {\bibinfo {author} {\bibfnamefont {M.}~\bibnamefont
  {Sigrist}}\ and\ \bibinfo {author} {\bibfnamefont {K.}~\bibnamefont {Ueda}},\
  }\bibfield  {title} {\bibinfo {title} {Phenomenological theory of
  unconventional superconductivity},\ }\href
  {https://doi.org/10.1103/RevModPhys.63.239} {\bibfield  {journal} {\bibinfo
  {journal} {Rev. Mod. Phys.}\ }\textbf {\bibinfo {volume} {63}},\ \bibinfo
  {pages} {239} (\bibinfo {year} {1991})}\BibitemShut {NoStop}%
\bibitem [{\citenamefont {Hillier}\ \emph {et~al.}(2009)\citenamefont
  {Hillier}, \citenamefont {Quintanilla},\ and\ \citenamefont
  {Cywinski}}]{Hillier2009}%
  \BibitemOpen
  \bibfield  {author} {\bibinfo {author} {\bibfnamefont {A.~D.}\ \bibnamefont
  {Hillier}}, \bibinfo {author} {\bibfnamefont {J.}~\bibnamefont
  {Quintanilla}},\ and\ \bibinfo {author} {\bibfnamefont {R.}~\bibnamefont
  {Cywinski}},\ }\bibfield  {title} {\bibinfo {title} {Evidence for
  time-reversal symmetry breaking in the noncentrosymmetric superconductor
  {LaNiC$_2$}},\ }\href {https://doi.org/10.1103/PhysRevLett.102.117007}
  {\bibfield  {journal} {\bibinfo  {journal} {Phys. Rev. Lett.}\ }\textbf
  {\bibinfo {volume} {102}},\ \bibinfo {pages} {117007} (\bibinfo {year}
  {2009})}\BibitemShut {NoStop}%
\bibitem [{\citenamefont {Chen}\ \emph {et~al.}(2013)\citenamefont {Chen},
  \citenamefont {Jiao}, \citenamefont {Zhang}, \citenamefont {Chen},
  \citenamefont {Yang}, \citenamefont {Nicklas}, \citenamefont {Steglich},\
  and\ \citenamefont {Yuan}}]{chen2013}%
  \BibitemOpen
  \bibfield  {author} {\bibinfo {author} {\bibfnamefont {J.}~\bibnamefont
  {Chen}}, \bibinfo {author} {\bibfnamefont {L.}~\bibnamefont {Jiao}}, \bibinfo
  {author} {\bibfnamefont {J.}~\bibnamefont {Zhang}}, \bibinfo {author}
  {\bibfnamefont {Y.}~\bibnamefont {Chen}}, \bibinfo {author} {\bibfnamefont
  {L.}~\bibnamefont {Yang}}, \bibinfo {author} {\bibfnamefont {M.}~\bibnamefont
  {Nicklas}}, \bibinfo {author} {\bibfnamefont {F.}~\bibnamefont {Steglich}},\
  and\ \bibinfo {author} {\bibfnamefont {H.}~\bibnamefont {Yuan}},\ }\bibfield
  {title} {\bibinfo {title} {Evidence for two-gap superconductivity in the
  non-centrosymmetric compound {LaNiC$_2$}},\ }\href
  {https://doi.org/https://doi.org/10.1088/1367-2630/15/5/053005} {\bibfield
  {journal} {\bibinfo  {journal} {New J. Phys.}\ }\textbf {\bibinfo {volume}
  {15}},\ \bibinfo {pages} {053005} (\bibinfo {year} {2013})}\BibitemShut
  {NoStop}%
\bibitem [{\citenamefont {Hillier}\ \emph {et~al.}(2012)\citenamefont
  {Hillier}, \citenamefont {Quintanilla}, \citenamefont {Mazidian},
  \citenamefont {Annett},\ and\ \citenamefont {Cywinski}}]{Hillier2012}%
  \BibitemOpen
  \bibfield  {author} {\bibinfo {author} {\bibfnamefont {A.~D.}\ \bibnamefont
  {Hillier}}, \bibinfo {author} {\bibfnamefont {J.}~\bibnamefont
  {Quintanilla}}, \bibinfo {author} {\bibfnamefont {B.}~\bibnamefont
  {Mazidian}}, \bibinfo {author} {\bibfnamefont {J.~F.}\ \bibnamefont
  {Annett}},\ and\ \bibinfo {author} {\bibfnamefont {R.}~\bibnamefont
  {Cywinski}},\ }\bibfield  {title} {\bibinfo {title} {Nonunitary triplet
  pairing in the centrosymmetric superconductor {LaNiGa$_2$}},\ }\href
  {https://doi.org/10.1103/PhysRevLett.109.097001} {\bibfield  {journal}
  {\bibinfo  {journal} {Phys. Rev. Lett.}\ }\textbf {\bibinfo {volume} {109}},\
  \bibinfo {pages} {097001} (\bibinfo {year} {2012})}\BibitemShut {NoStop}%
\bibitem [{\citenamefont {Weng}\ \emph {et~al.}(2016)\citenamefont {Weng},
  \citenamefont {Zhang}, \citenamefont {Smidman}, \citenamefont {Shang},
  \citenamefont {Quintanilla}, \citenamefont {Annett}, \citenamefont {Nicklas},
  \citenamefont {Pang}, \citenamefont {Jiao}, \citenamefont {Jiang},
  \citenamefont {Chen}, \citenamefont {Steglich},\ and\ \citenamefont
  {Yuan}}]{Weng2016}%
  \BibitemOpen
  \bibfield  {author} {\bibinfo {author} {\bibfnamefont {Z.~F.}\ \bibnamefont
  {Weng}}, \bibinfo {author} {\bibfnamefont {J.~L.}\ \bibnamefont {Zhang}},
  \bibinfo {author} {\bibfnamefont {M.}~\bibnamefont {Smidman}}, \bibinfo
  {author} {\bibfnamefont {T.}~\bibnamefont {Shang}}, \bibinfo {author}
  {\bibfnamefont {J.}~\bibnamefont {Quintanilla}}, \bibinfo {author}
  {\bibfnamefont {J.~F.}\ \bibnamefont {Annett}}, \bibinfo {author}
  {\bibfnamefont {M.}~\bibnamefont {Nicklas}}, \bibinfo {author} {\bibfnamefont
  {G.~M.}\ \bibnamefont {Pang}}, \bibinfo {author} {\bibfnamefont
  {L.}~\bibnamefont {Jiao}}, \bibinfo {author} {\bibfnamefont {W.~B.}\
  \bibnamefont {Jiang}}, \bibinfo {author} {\bibfnamefont {Y.}~\bibnamefont
  {Chen}}, \bibinfo {author} {\bibfnamefont {F.}~\bibnamefont {Steglich}},\
  and\ \bibinfo {author} {\bibfnamefont {H.~Q.}\ \bibnamefont {Yuan}},\
  }\bibfield  {title} {\bibinfo {title} {Two-gap superconductivity in
  {LaNiGa$_2$} with nonunitary triplet pairing and even parity gap symmetry},\
  }\href {https://doi.org/10.1103/PhysRevLett.117.027001} {\bibfield  {journal}
  {\bibinfo  {journal} {Phys. Rev. Lett.}\ }\textbf {\bibinfo {volume} {117}},\
  \bibinfo {pages} {027001} (\bibinfo {year} {2016})}\BibitemShut {NoStop}%
\bibitem [{\citenamefont {Ghosh}\ \emph
  {et~al.}(2020{\natexlab{b}})\citenamefont {Ghosh}, \citenamefont {Csire},
  \citenamefont {Whittlesea}, \citenamefont {Annett}, \citenamefont {Gradhand},
  \citenamefont {\'Ujfalussy},\ and\ \citenamefont {Quintanilla}}]{Ghosh2020b}%
  \BibitemOpen
  \bibfield  {author} {\bibinfo {author} {\bibfnamefont {S.~K.}\ \bibnamefont
  {Ghosh}}, \bibinfo {author} {\bibfnamefont {G.}~\bibnamefont {Csire}},
  \bibinfo {author} {\bibfnamefont {P.}~\bibnamefont {Whittlesea}}, \bibinfo
  {author} {\bibfnamefont {J.~F.}\ \bibnamefont {Annett}}, \bibinfo {author}
  {\bibfnamefont {M.}~\bibnamefont {Gradhand}}, \bibinfo {author}
  {\bibfnamefont {B.}~\bibnamefont {\'Ujfalussy}},\ and\ \bibinfo {author}
  {\bibfnamefont {J.}~\bibnamefont {Quintanilla}},\ }\bibfield  {title}
  {\bibinfo {title} {Quantitative theory of triplet pairing in the
  unconventional superconductor {LaNiGa$_2$}},\ }\href
  {https://doi.org/10.1103/PhysRevB.101.100506} {\bibfield  {journal} {\bibinfo
   {journal} {Phys. Rev. B}\ }\textbf {\bibinfo {volume} {101}},\ \bibinfo
  {pages} {100506} (\bibinfo {year} {2020}{\natexlab{b}})}\BibitemShut
  {NoStop}%
\bibitem [{\citenamefont {Badger}\ \emph {et~al.}(2021)\citenamefont {Badger},
  \citenamefont {Quan}, \citenamefont {Staab}, \citenamefont {Sumita},
  \citenamefont {Rossi}, \citenamefont {Devlin}, \citenamefont {Neubauer},
  \citenamefont {Shulman}, \citenamefont {Fettinger}, \citenamefont {Klavins}
  \emph {et~al.}}]{Badger2021}%
  \BibitemOpen
  \bibfield  {author} {\bibinfo {author} {\bibfnamefont {J.~R.}\ \bibnamefont
  {Badger}}, \bibinfo {author} {\bibfnamefont {Y.}~\bibnamefont {Quan}},
  \bibinfo {author} {\bibfnamefont {M.~C.}\ \bibnamefont {Staab}}, \bibinfo
  {author} {\bibfnamefont {S.}~\bibnamefont {Sumita}}, \bibinfo {author}
  {\bibfnamefont {A.}~\bibnamefont {Rossi}}, \bibinfo {author} {\bibfnamefont
  {K.~P.}\ \bibnamefont {Devlin}}, \bibinfo {author} {\bibfnamefont
  {K.}~\bibnamefont {Neubauer}}, \bibinfo {author} {\bibfnamefont {D.~S.}\
  \bibnamefont {Shulman}}, \bibinfo {author} {\bibfnamefont {J.~C.}\
  \bibnamefont {Fettinger}}, \bibinfo {author} {\bibfnamefont {P.}~\bibnamefont
  {Klavins}}, \emph {et~al.},\ }\bibfield  {title} {\bibinfo {title} {Dirac
  lines and loop at the \textsc{F}ermi level in the time-reversal symmetry
  breaking superconductor {LaNiGa$_2$}},\ }\href
  {https://arxiv.org/abs/2109.06983} {\bibfield  {journal} {\bibinfo  {journal}
  {arXiv preprint arXiv:2109.06983}\ } (\bibinfo {year} {2021})}\BibitemShut
  {NoStop}%
\bibitem [{\citenamefont {Ran}\ \emph {et~al.}(2019)\citenamefont {Ran},
  \citenamefont {Eckberg}, \citenamefont {Ding}, \citenamefont {Furukawa},
  \citenamefont {Metz}, \citenamefont {Saha}, \citenamefont {Liu},
  \citenamefont {Zic}, \citenamefont {Kim}, \citenamefont {Paglione} \emph
  {et~al.}}]{ran2019}%
  \BibitemOpen
  \bibfield  {author} {\bibinfo {author} {\bibfnamefont {S.}~\bibnamefont
  {Ran}}, \bibinfo {author} {\bibfnamefont {C.}~\bibnamefont {Eckberg}},
  \bibinfo {author} {\bibfnamefont {Q.-P.}\ \bibnamefont {Ding}}, \bibinfo
  {author} {\bibfnamefont {Y.}~\bibnamefont {Furukawa}}, \bibinfo {author}
  {\bibfnamefont {T.}~\bibnamefont {Metz}}, \bibinfo {author} {\bibfnamefont
  {S.~R.}\ \bibnamefont {Saha}}, \bibinfo {author} {\bibfnamefont {I.-L.}\
  \bibnamefont {Liu}}, \bibinfo {author} {\bibfnamefont {M.}~\bibnamefont
  {Zic}}, \bibinfo {author} {\bibfnamefont {H.}~\bibnamefont {Kim}}, \bibinfo
  {author} {\bibfnamefont {J.}~\bibnamefont {Paglione}}, \emph {et~al.},\
  }\bibfield  {title} {\bibinfo {title} {Nearly ferromagnetic spin-triplet
  superconductivity},\ }\href {https://doi.org/10.1126/science.aav8645}
  {\bibfield  {journal} {\bibinfo  {journal} {Science}\ }\textbf {\bibinfo
  {volume} {365}},\ \bibinfo {pages} {684} (\bibinfo {year}
  {2019})}\BibitemShut {NoStop}%
\bibitem [{\citenamefont {Metz}\ \emph {et~al.}(2019)\citenamefont {Metz},
  \citenamefont {Bae}, \citenamefont {Ran}, \citenamefont {Liu}, \citenamefont
  {Eo}, \citenamefont {Fuhrman}, \citenamefont {Agterberg}, \citenamefont
  {Anlage}, \citenamefont {Butch},\ and\ \citenamefont
  {Paglione}}]{Tristin2019}%
  \BibitemOpen
  \bibfield  {author} {\bibinfo {author} {\bibfnamefont {T.}~\bibnamefont
  {Metz}}, \bibinfo {author} {\bibfnamefont {S.}~\bibnamefont {Bae}}, \bibinfo
  {author} {\bibfnamefont {S.}~\bibnamefont {Ran}}, \bibinfo {author}
  {\bibfnamefont {I.-L.}\ \bibnamefont {Liu}}, \bibinfo {author} {\bibfnamefont
  {Y.~S.}\ \bibnamefont {Eo}}, \bibinfo {author} {\bibfnamefont {W.~T.}\
  \bibnamefont {Fuhrman}}, \bibinfo {author} {\bibfnamefont {D.~F.}\
  \bibnamefont {Agterberg}}, \bibinfo {author} {\bibfnamefont {S.~M.}\
  \bibnamefont {Anlage}}, \bibinfo {author} {\bibfnamefont {N.~P.}\
  \bibnamefont {Butch}},\ and\ \bibinfo {author} {\bibfnamefont
  {J.}~\bibnamefont {Paglione}},\ }\bibfield  {title} {\bibinfo {title}
  {Point-node gap structure of the spin-triplet superconductor {UTe$_2$}},\
  }\href {https://doi.org/10.1103/PhysRevB.100.220504} {\bibfield  {journal}
  {\bibinfo  {journal} {Phys. Rev. B}\ }\textbf {\bibinfo {volume} {100}},\
  \bibinfo {pages} {220504} (\bibinfo {year} {2019})}\BibitemShut {NoStop}%
\bibitem [{\citenamefont {Benndorf}\ \emph {et~al.}(2017)\citenamefont
  {Benndorf}, \citenamefont {Heletta}, \citenamefont {Heymann}, \citenamefont
  {Huppertz}, \citenamefont {Eckert},\ and\ \citenamefont
  {P{\"o}ttgen}}]{Benndorf2017}%
  \BibitemOpen
  \bibfield  {author} {\bibinfo {author} {\bibfnamefont {C.}~\bibnamefont
  {Benndorf}}, \bibinfo {author} {\bibfnamefont {L.}~\bibnamefont {Heletta}},
  \bibinfo {author} {\bibfnamefont {G.}~\bibnamefont {Heymann}}, \bibinfo
  {author} {\bibfnamefont {H.}~\bibnamefont {Huppertz}}, \bibinfo {author}
  {\bibfnamefont {H.}~\bibnamefont {Eckert}},\ and\ \bibinfo {author}
  {\bibfnamefont {R.}~\bibnamefont {P{\"o}ttgen}},\ }\bibfield  {title}
  {\bibinfo {title} {{NbOsSi} and {TaOsSi}--two new superconducting ternary
  osmium silicides},\ }\href
  {https://doi.org/10.1016/j.solidstatesciences.2017.04.002} {\bibfield
  {journal} {\bibinfo  {journal} {Solid State Sciences}\ }\textbf {\bibinfo
  {volume} {68}},\ \bibinfo {pages} {32} (\bibinfo {year} {2017})}\BibitemShut
  {NoStop}%
\bibitem [{\citenamefont {Haque}\ and\ \citenamefont
  {Hossain}(2018)}]{Haque2018}%
  \BibitemOpen
  \bibfield  {author} {\bibinfo {author} {\bibfnamefont {E.}~\bibnamefont
  {Haque}}\ and\ \bibinfo {author} {\bibfnamefont {M.~A.}\ \bibnamefont
  {Hossain}},\ }\bibfield  {title} {\bibinfo {title} {Elastic, electronic,
  thermodynamic and transport properties of {XOsSi} ({X= Nb, Ta})
  superconductors: First-principles calculations},\ }\href
  {https://doi.org/10.1016/j.jallcom.2017.12.300} {\bibfield  {journal}
  {\bibinfo  {journal} {J. Alloys and Compounds}\ }\textbf {\bibinfo {volume}
  {739}},\ \bibinfo {pages} {737} (\bibinfo {year} {2018})}\BibitemShut
  {NoStop}%
\bibitem [{SM()}]{SM}%
  \BibitemOpen
  \href@noop {} {}\bibinfo {note} {See the Supplemental Material at xxxx for
  details of the measurements of the crystal structure, heat capacity and
  critical field, as well as for the data analysis, DFT calculation, symmetry
  analysis and toy-model calculations.}\BibitemShut {Stop}%
\bibitem [{\citenamefont {Kubo}(1981)}]{Kubo1981}%
  \BibitemOpen
  \bibfield  {author} {\bibinfo {author} {\bibfnamefont {R.}~\bibnamefont
  {Kubo}},\ }\bibfield  {title} {\bibinfo {title} {A stochastic theory of spin
  relaxation},\ }\href {https://doi.org/10.1007/BF01037553} {\bibfield
  {journal} {\bibinfo  {journal} {Hyperfine Interactions}\ }\textbf {\bibinfo
  {volume} {8}},\ \bibinfo {pages} {731} (\bibinfo {year} {1981})}\BibitemShut
  {NoStop}%
\bibitem [{\citenamefont {Sonier}\ \emph {et~al.}(2000)\citenamefont {Sonier},
  \citenamefont {Brewer},\ and\ \citenamefont {Kiefl}}]{Sonier2000}%
  \BibitemOpen
  \bibfield  {author} {\bibinfo {author} {\bibfnamefont {J.~E.}\ \bibnamefont
  {Sonier}}, \bibinfo {author} {\bibfnamefont {J.~H.}\ \bibnamefont {Brewer}},\
  and\ \bibinfo {author} {\bibfnamefont {R.~F.}\ \bibnamefont {Kiefl}},\
  }\bibfield  {title} {\bibinfo {title} {{$\mu$SR studies of the vortex state
  in type-II superconductors}},\ }\href
  {https://doi.org/10.1103/revmodphys.72.769} {\bibfield  {journal} {\bibinfo
  {journal} {Rev. Mod. Phys.}\ }\textbf {\bibinfo {volume} {72}},\ \bibinfo
  {pages} {769} (\bibinfo {year} {2000})}\BibitemShut {NoStop}%
\bibitem [{\citenamefont {Brandt}(2003)}]{Brandt2003}%
  \BibitemOpen
  \bibfield  {author} {\bibinfo {author} {\bibfnamefont {E.~H.}\ \bibnamefont
  {Brandt}},\ }\bibfield  {title} {\bibinfo {title} {Properties of the ideal
  {Ginzburg-Landau} vortex lattice},\ }\href
  {https://doi.org/10.1103/physrevb.68.054506} {\bibfield  {journal} {\bibinfo
  {journal} {Phys. Rev. B}\ }\textbf {\bibinfo {volume} {68}},\ \bibinfo
  {pages} {054506} (\bibinfo {year} {2003})}\BibitemShut {NoStop}%
\bibitem [{\citenamefont {Prozorov}\ and\ \citenamefont
  {Giannetta}(2006)}]{Prozorov2006}%
  \BibitemOpen
  \bibfield  {author} {\bibinfo {author} {\bibfnamefont {R.}~\bibnamefont
  {Prozorov}}\ and\ \bibinfo {author} {\bibfnamefont {R.~W.}\ \bibnamefont
  {Giannetta}},\ }\bibfield  {title} {\bibinfo {title} {Magnetic penetration
  depth in unconventional superconductors},\ }\href
  {https://doi.org/10.1002/chin.200702273} {\bibfield  {journal} {\bibinfo
  {journal} {Supercond. Sci. Tech.}\ }\textbf {\bibinfo {volume} {19}},\
  \bibinfo {pages} {R41} (\bibinfo {year} {2006})}\BibitemShut {NoStop}%
\bibitem [{\citenamefont {Carrington}\ and\ \citenamefont
  {Manzano}(2003)}]{Carrington2003}%
  \BibitemOpen
  \bibfield  {author} {\bibinfo {author} {\bibfnamefont {A.}~\bibnamefont
  {Carrington}}\ and\ \bibinfo {author} {\bibfnamefont {F.}~\bibnamefont
  {Manzano}},\ }\bibfield  {title} {\bibinfo {title} {Magnetic penetration
  depth of {MgB$_2$}},\ }\href {https://doi.org/10.1016/S0921-4534(02)02319-5}
  {\bibfield  {journal} {\bibinfo  {journal} {Physica C}\ }\textbf {\bibinfo
  {volume} {385}},\ \bibinfo {pages} {205} (\bibinfo {year}
  {2003})}\BibitemShut {NoStop}%
\bibitem [{\citenamefont {Xu}\ \emph {et~al.}(2019)\citenamefont {Xu},
  \citenamefont {Li}, \citenamefont {Feng}, \citenamefont {Jiao}, \citenamefont
  {Li}, \citenamefont {Liu}, \citenamefont {Zhou}, \citenamefont {Sankar},
  \citenamefont {Zhigadlo}, \citenamefont {Wang}, \citenamefont {Han},
  \citenamefont {Qian}, \citenamefont {Ye}, \citenamefont {Zhou}, \citenamefont
  {Shiroka}, \citenamefont {Biswas}, \citenamefont {Xu},\ and\ \citenamefont
  {Shi}}]{Xu2019}%
  \BibitemOpen
  \bibfield  {author} {\bibinfo {author} {\bibfnamefont {C.~Q.}\ \bibnamefont
  {Xu}}, \bibinfo {author} {\bibfnamefont {B.}~\bibnamefont {Li}}, \bibinfo
  {author} {\bibfnamefont {J.~J.}\ \bibnamefont {Feng}}, \bibinfo {author}
  {\bibfnamefont {W.~H.}\ \bibnamefont {Jiao}}, \bibinfo {author}
  {\bibfnamefont {Y.~K.}\ \bibnamefont {Li}}, \bibinfo {author} {\bibfnamefont
  {S.~W.}\ \bibnamefont {Liu}}, \bibinfo {author} {\bibfnamefont {Y.~X.}\
  \bibnamefont {Zhou}}, \bibinfo {author} {\bibfnamefont {R.}~\bibnamefont
  {Sankar}}, \bibinfo {author} {\bibfnamefont {N.~D.}\ \bibnamefont
  {Zhigadlo}}, \bibinfo {author} {\bibfnamefont {H.~B.}\ \bibnamefont {Wang}},
  \bibinfo {author} {\bibfnamefont {Z.~D.}\ \bibnamefont {Han}}, \bibinfo
  {author} {\bibfnamefont {B.}~\bibnamefont {Qian}}, \bibinfo {author}
  {\bibfnamefont {W.}~\bibnamefont {Ye}}, \bibinfo {author} {\bibfnamefont
  {W.}~\bibnamefont {Zhou}}, \bibinfo {author} {\bibfnamefont {T.}~\bibnamefont
  {Shiroka}}, \bibinfo {author} {\bibfnamefont {P.~K.}\ \bibnamefont {Biswas}},
  \bibinfo {author} {\bibfnamefont {X.}~\bibnamefont {Xu}},\ and\ \bibinfo
  {author} {\bibfnamefont {Z.~X.}\ \bibnamefont {Shi}},\ }\bibfield  {title}
  {\bibinfo {title} {Two-gap superconductivity and topological surface states
  in {TaOsSi}},\ }\href {https://doi.org/10.1103/PhysRevB.100.134503}
  {\bibfield  {journal} {\bibinfo  {journal} {Phys. Rev. B}\ }\textbf {\bibinfo
  {volume} {100}},\ \bibinfo {pages} {134503} (\bibinfo {year}
  {2019})}\BibitemShut {NoStop}%
\bibitem [{non()}]{nonsymmnote}%
  \BibitemOpen
  \href@noop {} {\emph {\bibinfo {title} {{\rm The nonsymmorphic symmetries
  present in {(Ta, Nb)OsSi}, in general, can give rise to superconducting order
  parameters with additional symmetry required nodes along the high symmetry
  directions in the Brillouin zone boundaries but cannot give rise to a
  multicomponent order parameter to facilitate TRS
  breaking~\cite{sumita2018,sumita2019,Badger2021}.}}}}\BibitemShut {Stop}%
\bibitem [{\citenamefont {Araki}\ \emph {et~al.}(2019)\citenamefont {Araki},
  \citenamefont {Kohei}, \citenamefont {Tanaka}, \citenamefont {Nakamura},
  \citenamefont {Nojima}, \citenamefont {Ochiai},\ and\ \citenamefont
  {Katoh}}]{Araki2019}%
  \BibitemOpen
  \bibfield  {author} {\bibinfo {author} {\bibfnamefont {K.}~\bibnamefont
  {Araki}}, \bibinfo {author} {\bibfnamefont {T.}~\bibnamefont {Kohei}},
  \bibinfo {author} {\bibfnamefont {H.}~\bibnamefont {Tanaka}}, \bibinfo
  {author} {\bibfnamefont {S.}~\bibnamefont {Nakamura}}, \bibinfo {author}
  {\bibfnamefont {T.}~\bibnamefont {Nojima}}, \bibinfo {author} {\bibfnamefont
  {A.}~\bibnamefont {Ochiai}},\ and\ \bibinfo {author} {\bibfnamefont
  {K.}~\bibnamefont {Katoh}},\ }\bibfield  {title} {\bibinfo {title} {Magnetic
  and transport properties of {YbNiGe} with a {TiNiSi}-type structure},\ }\href
  {https://doi.org/10.7566/JPSJ.88.114709} {\bibfield  {journal} {\bibinfo
  {journal} {J. Phys. Soc. Jpn.}\ }\textbf {\bibinfo {volume} {88}},\ \bibinfo
  {pages} {114709} (\bibinfo {year} {2019})}\BibitemShut {NoStop}%
\bibitem [{\citenamefont {Wu}\ \emph {et~al.}(2017)\citenamefont {Wu},
  \citenamefont {Sumida}, \citenamefont {Miyamoto}, \citenamefont {Taguchi},
  \citenamefont {Yoshikawa}, \citenamefont {Kimura}, \citenamefont {Ueda},
  \citenamefont {Arita}, \citenamefont {Nagao}, \citenamefont {Watauchi} \emph
  {et~al.}}]{Wu2017}%
  \BibitemOpen
  \bibfield  {author} {\bibinfo {author} {\bibfnamefont {S.-L.}\ \bibnamefont
  {Wu}}, \bibinfo {author} {\bibfnamefont {K.}~\bibnamefont {Sumida}}, \bibinfo
  {author} {\bibfnamefont {K.}~\bibnamefont {Miyamoto}}, \bibinfo {author}
  {\bibfnamefont {K.}~\bibnamefont {Taguchi}}, \bibinfo {author} {\bibfnamefont
  {T.}~\bibnamefont {Yoshikawa}}, \bibinfo {author} {\bibfnamefont
  {A.}~\bibnamefont {Kimura}}, \bibinfo {author} {\bibfnamefont
  {Y.}~\bibnamefont {Ueda}}, \bibinfo {author} {\bibfnamefont {M.}~\bibnamefont
  {Arita}}, \bibinfo {author} {\bibfnamefont {M.}~\bibnamefont {Nagao}},
  \bibinfo {author} {\bibfnamefont {S.}~\bibnamefont {Watauchi}}, \emph
  {et~al.},\ }\bibfield  {title} {\bibinfo {title} {Direct evidence of hidden
  local spin polarization in a centrosymmetric superconductor
  {LaO$_{0.55}$F$_{0.45}$BiS$_2$}},\ }\href
  {https://doi.org/10.1038/s41467-017-02058-2} {\bibfield  {journal} {\bibinfo
  {journal} {Nat. Commun.}\ }\textbf {\bibinfo {volume} {8}},\ \bibinfo {pages}
  {1} (\bibinfo {year} {2017})}\BibitemShut {NoStop}%
\bibitem [{\citenamefont {Zhang}\ \emph {et~al.}(2014)\citenamefont {Zhang},
  \citenamefont {Liu}, \citenamefont {Luo}, \citenamefont {Freeman},\ and\
  \citenamefont {Zunger}}]{Zhang2014}%
  \BibitemOpen
  \bibfield  {author} {\bibinfo {author} {\bibfnamefont {X.}~\bibnamefont
  {Zhang}}, \bibinfo {author} {\bibfnamefont {Q.}~\bibnamefont {Liu}}, \bibinfo
  {author} {\bibfnamefont {J.-W.}\ \bibnamefont {Luo}}, \bibinfo {author}
  {\bibfnamefont {A.~J.}\ \bibnamefont {Freeman}},\ and\ \bibinfo {author}
  {\bibfnamefont {A.}~\bibnamefont {Zunger}},\ }\bibfield  {title} {\bibinfo
  {title} {Hidden spin polarization in inversion-symmetric bulk crystals},\
  }\href {https://doi.org/10.1038/nphys2933} {\bibfield  {journal} {\bibinfo
  {journal} {Nat. Phys.}\ }\textbf {\bibinfo {volume} {10}},\ \bibinfo {pages}
  {387} (\bibinfo {year} {2014})}\BibitemShut {NoStop}%
\bibitem [{old()}]{oldspheat}%
  \BibitemOpen
  \href@noop {} {}\bibinfo {note} {Although the experimental specific heat data
  for TaOsSi was shown to be fitted well by a two-gap model having more fitting
  parameters~\cite{Xu2019}, we note from \fig{fig:Cv_fig} that the two-band toy
  model in the INT ground state giving rise to a single full gap provides a
  very good fitting as well.}\BibitemShut {Stop}%
\bibitem [{\citenamefont {Hayes}\ \emph {et~al.}(2021)\citenamefont {Hayes},
  \citenamefont {Wei}, \citenamefont {Metz}, \citenamefont {Zhang},
  \citenamefont {Eo}, \citenamefont {Ran}, \citenamefont {Saha}, \citenamefont
  {Collini}, \citenamefont {Butch}, \citenamefont {Agterberg} \emph
  {et~al.}}]{hayes2021}%
  \BibitemOpen
  \bibfield  {author} {\bibinfo {author} {\bibfnamefont {I.}~\bibnamefont
  {Hayes}}, \bibinfo {author} {\bibfnamefont {D.}~\bibnamefont {Wei}}, \bibinfo
  {author} {\bibfnamefont {T.}~\bibnamefont {Metz}}, \bibinfo {author}
  {\bibfnamefont {J.}~\bibnamefont {Zhang}}, \bibinfo {author} {\bibfnamefont
  {Y.}~\bibnamefont {Eo}}, \bibinfo {author} {\bibfnamefont {S.}~\bibnamefont
  {Ran}}, \bibinfo {author} {\bibfnamefont {S.}~\bibnamefont {Saha}}, \bibinfo
  {author} {\bibfnamefont {J.}~\bibnamefont {Collini}}, \bibinfo {author}
  {\bibfnamefont {N.}~\bibnamefont {Butch}}, \bibinfo {author} {\bibfnamefont
  {D.}~\bibnamefont {Agterberg}}, \emph {et~al.},\ }\bibfield  {title}
  {\bibinfo {title} {Multicomponent superconducting order parameter in
  {UTe$_2$}},\ }\href {https://doi.org/10.1126/science.abb0272} {\bibfield
  {journal} {\bibinfo  {journal} {Science}\ }\textbf {\bibinfo {volume}
  {373}},\ \bibinfo {pages} {797} (\bibinfo {year} {2021})}\BibitemShut
  {NoStop}%
\bibitem [{\citenamefont {Zhong}\ \emph {et~al.}(1986)\citenamefont {Zhong},
  \citenamefont {Chevalier}, \citenamefont {Etourneau},\ and\ \citenamefont
  {Hagenmuller}}]{Zhong1986}%
  \BibitemOpen
  \bibfield  {author} {\bibinfo {author} {\bibfnamefont {W.~X.}\ \bibnamefont
  {Zhong}}, \bibinfo {author} {\bibfnamefont {B.}~\bibnamefont {Chevalier}},
  \bibinfo {author} {\bibfnamefont {J.}~\bibnamefont {Etourneau}},\ and\
  \bibinfo {author} {\bibfnamefont {P.}~\bibnamefont {Hagenmuller}},\
  }\bibfield  {title} {\bibinfo {title} {Relationships between occurrence of
  superconductivity and crystal structure in new equiatomic ternary silicides
  {MTSi (M = Ti, Zr, Hf and T = Ru, Os, Rh)}},\ }\href
  {https://doi.org/https://doi.org/10.1016/0038-1098(86)90640-X} {\bibfield
  {journal} {\bibinfo  {journal} {Solid State Communications}\ }\textbf
  {\bibinfo {volume} {59}},\ \bibinfo {pages} {839} (\bibinfo {year}
  {1986})}\BibitemShut {NoStop}%
\bibitem [{\citenamefont {Sumita}\ and\ \citenamefont
  {Yanase}(2018)}]{sumita2018}%
  \BibitemOpen
  \bibfield  {author} {\bibinfo {author} {\bibfnamefont {S.}~\bibnamefont
  {Sumita}}\ and\ \bibinfo {author} {\bibfnamefont {Y.}~\bibnamefont
  {Yanase}},\ }\bibfield  {title} {\bibinfo {title} {Unconventional
  superconducting gap structure protected by space group symmetry},\ }\href
  {https://doi.org/10.1103/PhysRevB.97.134512} {\bibfield  {journal} {\bibinfo
  {journal} {Phys. Rev. B}\ }\textbf {\bibinfo {volume} {97}},\ \bibinfo
  {pages} {134512} (\bibinfo {year} {2018})}\BibitemShut {NoStop}%
\bibitem [{\citenamefont {Sumita}\ \emph {et~al.}(2019)\citenamefont {Sumita},
  \citenamefont {Nomoto}, \citenamefont {Shiozaki},\ and\ \citenamefont
  {Yanase}}]{sumita2019}%
  \BibitemOpen
  \bibfield  {author} {\bibinfo {author} {\bibfnamefont {S.}~\bibnamefont
  {Sumita}}, \bibinfo {author} {\bibfnamefont {T.}~\bibnamefont {Nomoto}},
  \bibinfo {author} {\bibfnamefont {K.}~\bibnamefont {Shiozaki}},\ and\
  \bibinfo {author} {\bibfnamefont {Y.}~\bibnamefont {Yanase}},\ }\bibfield
  {title} {\bibinfo {title} {Classification of topological crystalline
  superconducting nodes on high-symmetry lines: Point nodes, line nodes, and
  \textsc{B}ogoliubov \textsc{F}ermi surfaces},\ }\href
  {https://doi.org/10.1103/PhysRevB.99.134513} {\bibfield  {journal} {\bibinfo
  {journal} {Phys. Rev. B}\ }\textbf {\bibinfo {volume} {99}},\ \bibinfo
  {pages} {134513} (\bibinfo {year} {2019})}\BibitemShut {NoStop}%
\bibitem [{\citenamefont {Tinkham}(1996)}]{Tinkham1996}%
  \BibitemOpen
  \bibfield  {author} {\bibinfo {author} {\bibfnamefont {M.}~\bibnamefont
  {Tinkham}},\ }\href@noop {} {\emph {\bibinfo {title} {Introduction to
  Superconductivity}}}\ (\bibinfo  {publisher} {McGraw-Hill Inc.},\ \bibinfo
  {year} {1996})\BibitemShut {NoStop}%
\end{thebibliography}%


\newwrite\tempfile
\immediate\openout\tempfile=junkSM.\jobname
\immediate\write\tempfile{\noexpand{\thepage} }
\immediate\closeout\tempfile

\clearpage

\newpage

\appendix

\renewcommand{\appendixname}{}
\renewcommand{\thesection}{{S\arabic{section}}}
\renewcommand{\theequation}{S.\arabic{equation}}
\renewcommand{\thefigure}{S.\arabic{figure}}

\setcounter{page}{1}
\setcounter{figure}{0}
\setcounter{equation}{0}

\widetext

\centerline{\bf Supplemental Material}
\centerline{\bf for}
\centerline{\bf ``\titlename''}

\centerline{}

Abstract: In this Supplemental Material, we present details of the synthesis, characterization measurements, experimental methods and data analysis for the (Ta, Nb)OsSi materials. We also give additional band structure results, details of the symmetry analysis and computation of the specific heat using the Bogoliubov de-Gennes formalism for a toy model in the INT state.


\section{Synthesis and characterization of (Ta, Nb)OsSi samples}

\begin{figure}[htb]
\begin{center}
\includegraphics[width=\textwidth]{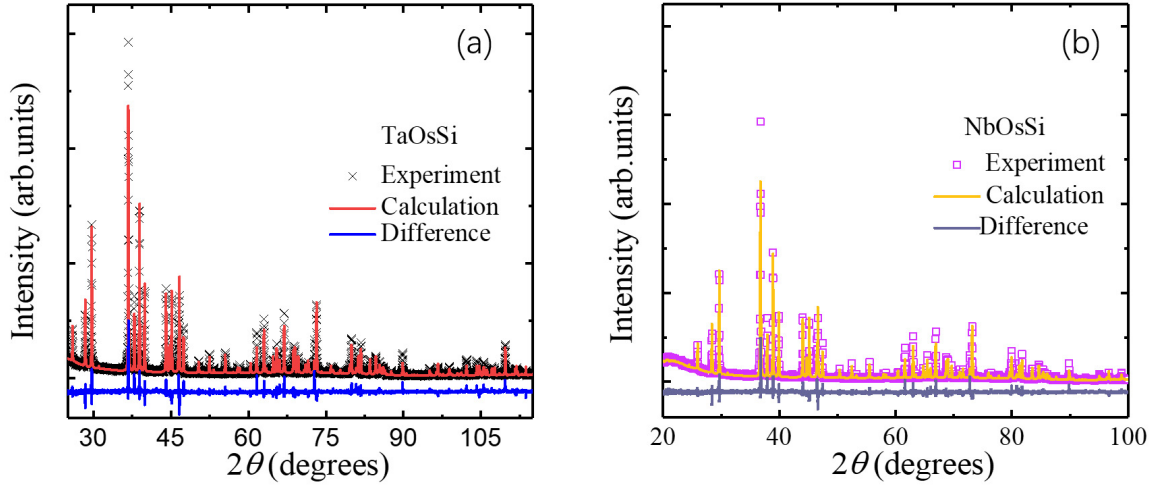}
\vspace{-0.25cm}
\caption{\textbf{XRD pattern.} Powder XRD pattern and Rietveld refinement for (a) TaOsSi and (b) NbOsSi.}
\label{fig:PXRD}
\end{center}
\end{figure}

\begin{figure}[htb]
\begin{center}
\includegraphics[width=\textwidth]{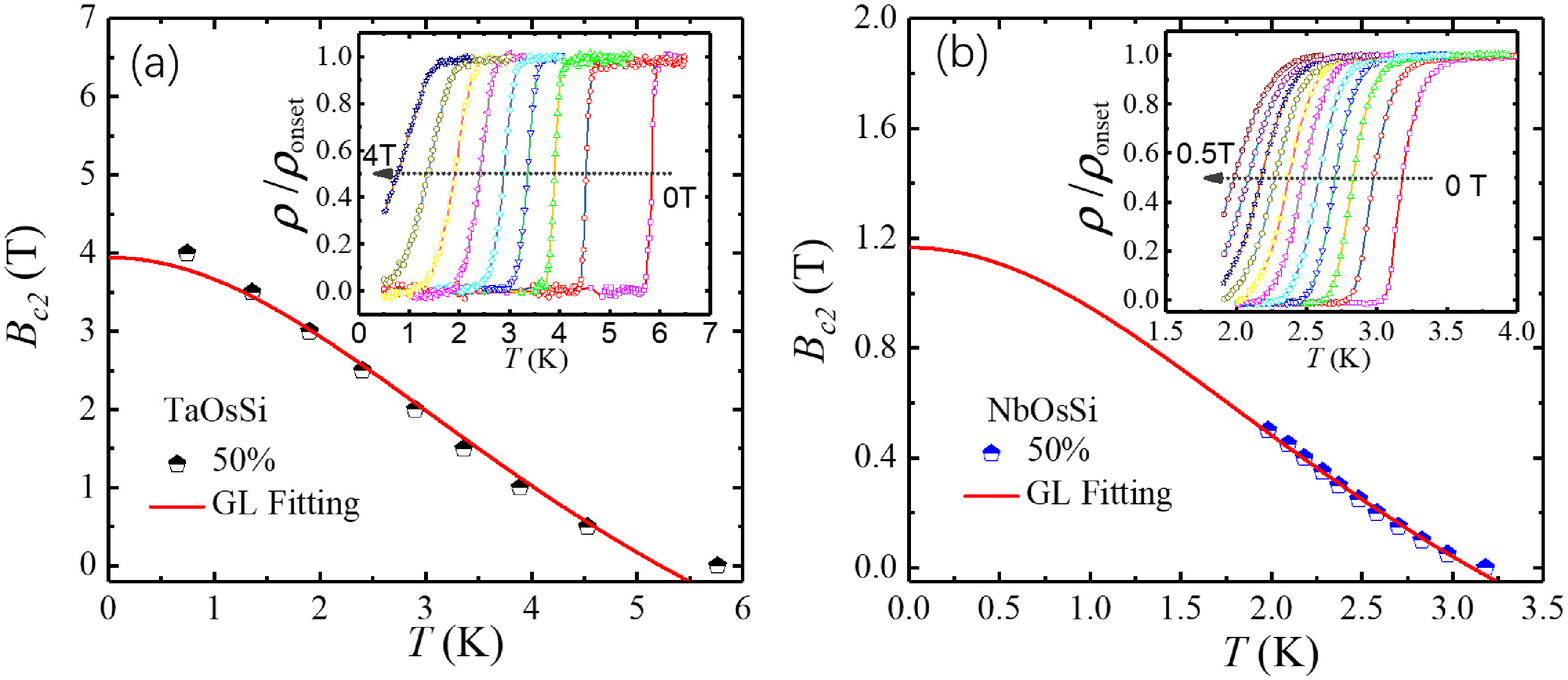}
\vspace{-0.25cm}
\caption{\textbf{Upper critical field.} Temperature dependence of the upper critical field $B_{c2}$, as determined by the $50\%$ criteria at which the resistivity drops to $50\%$ of its normal-state value $\rho_{onset}$ just above $T_c$ (illustrated in the insets) for (a) TaOsSi and (b) NbOsSi. Red solid lines represent the fits based on the Ginzburg–Landau (GL) theory.}
\label{fig:crit}
\end{center}
\end{figure}

Polycrystalline TaOsSi and NbOsSi were synthesized by the arc-melting method~\cite{Benndorf2017,Xu2019}. The room-temperature crystal structure was characterized by X-ray diffraction (XRD) equipped in a Rigaku diffractometer with Cu K-$\alpha$ radiation and a graphite monochromator (see \fig{fig:PXRD}). The crystallographic parameters are listed in \tab{tab:structure} for concreteness. Resistivity (the upper critical field, see \fig{fig:crit}), heat capacity properties were measured using Quantum Design Physical Property Measurement System (QD-PPMS). The magnetic susceptibility was studied in Quantum Design Magnetic Property Measurement System (QD-MPMS). We note from \fig{fig:crit} that the zero temperature values of the upper critical fields for TaOsSi and NbOsSi are $\approx 4$ T and $1.2$ T respectively, both of which are much larger than the applied transverse fields in the TF-$\mu$SR measurements.

\begin{table}[h]
\caption{Crystallographic lattice parameters obtained for TaOsSi and NbOsSi.}
\begin{center}
\begin{tabular}{l l l l l l l l l l l l l}
 \hline
Material && \hspace{1cm} $a$(\AA) && \hspace{1cm} $b$(\AA) && \hspace{1cm} $c$(\AA)  \\
\hline
TaOsSi && \hspace{1cm} 6.26(5) && \hspace{1cm} 3.89(3) && \hspace{1cm} 7.25(9)    \\
NbOsSi && \hspace{1cm} 6.28(1) && \hspace{1cm} 3.89(7) && \hspace{1cm} 7.27(3)   \\
\hline \hline
\end{tabular}
\end{center}
\label{tab:structure}
\end{table}

\begin{table}[h]
\caption{Parameters related to the experimental specific heat data.}
\begin{center}
\begin{tabular}{l l l l l l l l l l l l l}
 \hline
Material && \hspace{1cm} $T_{\rm c}$ (K) && \hspace{1cm} $\gamma$ (${\rm mJ/mol-K}^2$) && \hspace{1cm} $\beta$ (${\rm mJ/mol-K}^4$) && \hspace{1cm} $\delta$ (${\rm mJ/mol-K}^6$)  \\
\hline
TaOsSi && \hspace{1cm} 5.6 && \hspace{1cm} 7.90 && \hspace{1cm} 0.044 && \hspace{1cm} $0.0014$   \\
NbOsSi && \hspace{1cm} 3.2 && \hspace{1cm} 14.15 && \hspace{1cm} 0.950  && \hspace{1cm} 0.0310  \\
\hline \hline
\end{tabular}
\end{center}
\label{tab:cv}
\end{table}

\begin{figure}[htb]
\centering
\includegraphics[width=0.95\columnwidth]{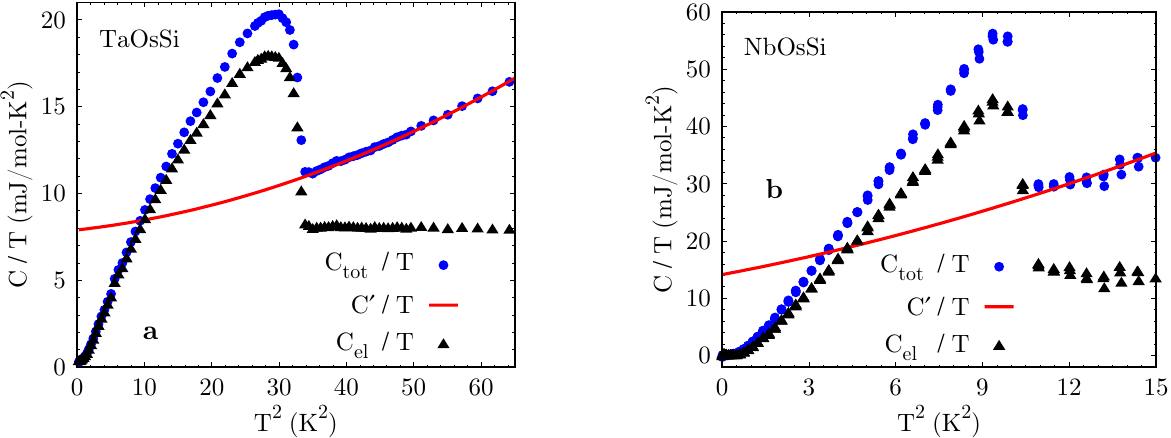}
\caption{\textbf{Specific heat.} Experimental specific heat data for TaOsSi in \textbf{a} and for NbOsSi in \textbf{b}.}
\label{fig:Cv_SM}
\end{figure}

The heat capacity was measured in zero field using a Quantum Design Physical Property Measurement System (QD-PPMS) with a $^3$He insert to get down to 0.5 K. The total specific heat $C_{tot}$  at low temperatures is made up of mainly two contributions,
\beq
C_{tot} =  C_{el} + C_{ph}
\eeq
where $C_{el}$ is the electronic specific heat having the form in the normal state
\beq
C_{el} = \gamma T
\eeq
with $\gamma$ being the Sommerfeld coefficient, $C_{ph}$ is the specific heat due to the phonons given by
\beq
C_{ph} = \beta T^3 + \delta T^5
\eeq
with $\beta$ and $\delta$ being temperature independent parameters. We define
\beq
C' = \gamma T + \beta T^3 + \delta T^5.
\eeq
We extract the electronic specific heat from the experimentally determined total specific heat by fitting the specific heat in the normal state with $C'$ as shown in the \fig{fig:Cv_SM} and the corresponding parameters are shown in the \tab{tab:cv}.

\section{Band structure}
The crystal structure of (Ta,Nb)OsSi is centrosymmetric and orthorhombic. The space group Pnma (no. 62) is nonsymmorphic which has three glide mirror planes: $G_1 = \{m_{(0,1,0)}|t_{(0,1/2,0)} \}$, $G_2 = \{m_{(0,0,1)}|t_{(1/2,0,1/2)} \}$ and $G_3 = \{m_{(1,0,0)}|t_{(1/2,1/2,1/2)} \}$ where $m_{(\alpha,\beta,\gamma)}$ denotes reflection in the plane $(\alpha,\beta,\gamma)$ and $t$ is a fractional translation parallel to the plane $(\alpha,\beta,\gamma)$. Then we can define the composite anti-unitary symmetry operations $\Theta_i = G_i \star \mathcal{T}$ with $i = 1$, 2 and 3, and $\mathcal{T}$ is the time-reversal operator, such that  
\bea
\Theta_1 (x,y,z,t) & \longrightarrow & (x,-y+\frac{1}{2},z,-t) \\
\Theta_2 (x,y,z,t) & \longrightarrow & (x+\frac{1}{2},y,-z+\frac{1}{2},-t) \\
\Theta_3 (x,y,z,t) & \longrightarrow & (-x+\frac{1}{2},y+\frac{1}{2},z+\frac{1}{2},-t) 
\eea
where $(x,y,z$) are the spatial coordinates along the $a$-, $b$- and $c$-axes in
units of the corresponding lattice constants. Most importantly, $\Theta_1^2 = 1$, $\Theta_2^2 = e^{-i k_x a}$ and $\Theta_3^2 = e^{-i (k_y b + k_z c)}$. Then at the Brillouin zone boundary $k_x a = \pi$ we have $\Theta_2^2 = -1$ which leads to two-fold degeneracy of all eigenstates in this plane.

\begin{figure}[htb]
\centering
\includegraphics[width=0.8\columnwidth]{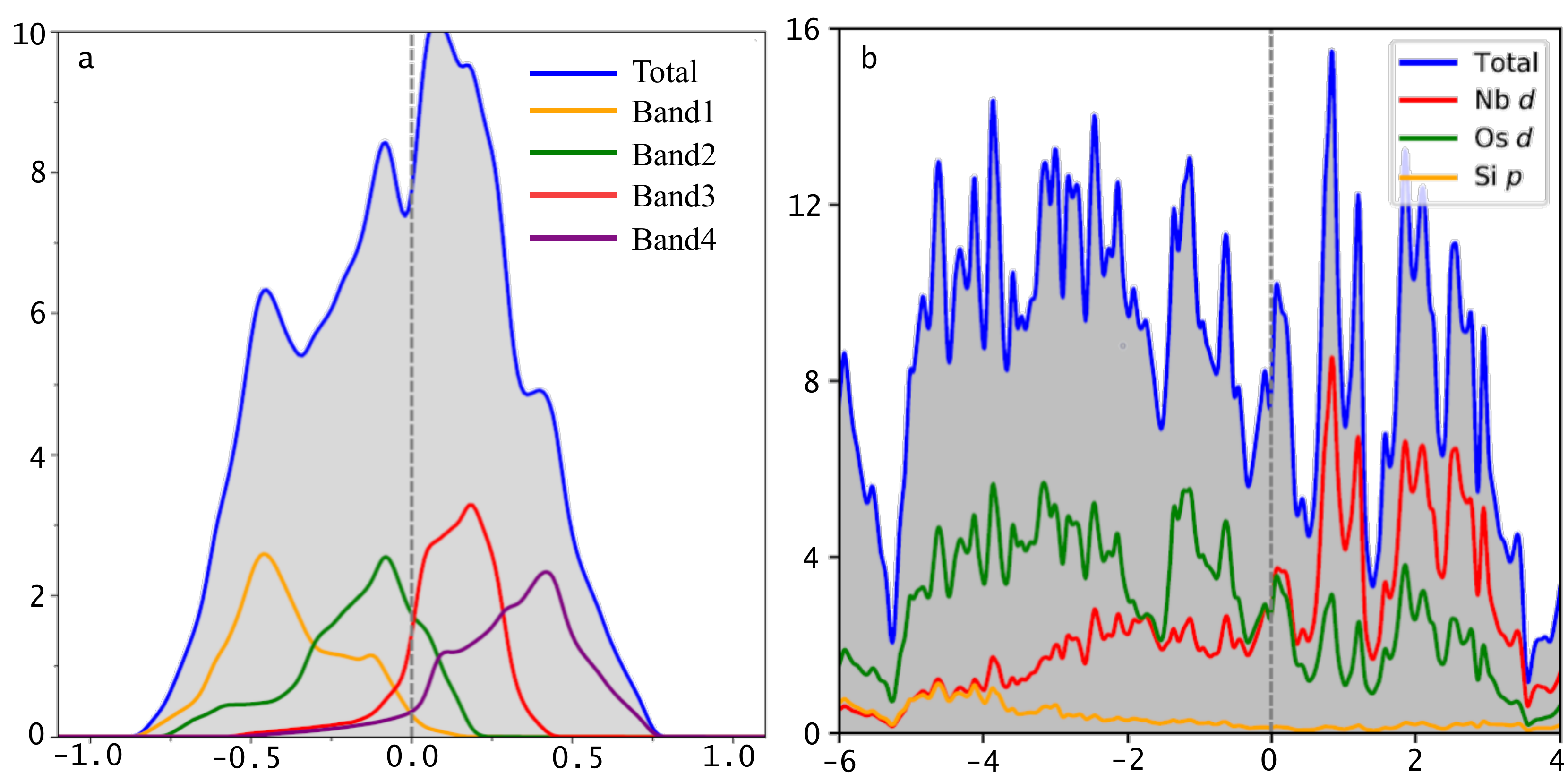}
\caption{\textbf{Projected density of states for NbOsSi.} \textbf{a} Contribution of the four Fermi surfaces without SOC to the density of states for NbOsSi. \textbf{b} Contribution of the different orbitals of the different atoms to the density of states for NbOsSi.}
\label{fig:pdos_SM}
\end{figure}

We have performed detailed band structure calculations of $A$OsSi ($A$ = Ta and Nb) using density functional theory (DFT) within the generalized gradient approximation (GGA). The four Fermi surface sheets of NbOsSi (without SOC) are shown in \fig{fig:FS}. The Fermi surface sheets shown in the panels (a), (b), (c) and (d) of \fig{fig:FS} contribute 8.1\%, 45.1\%, 37.6\% and 9.2\% respectively to the density of states at the Fermi level as seen from \fig{fig:pdos_SM}\textbf{a}. We note from \fig{fig:pdos_SM}\textbf{b} that the Nb 4$d$-orbitals and Os 5$d$-orbitals contribute the most to the density of states at the Fermi level.

\begin{figure*}[!hbt]
\centering
\includegraphics[width=0.99\textwidth]{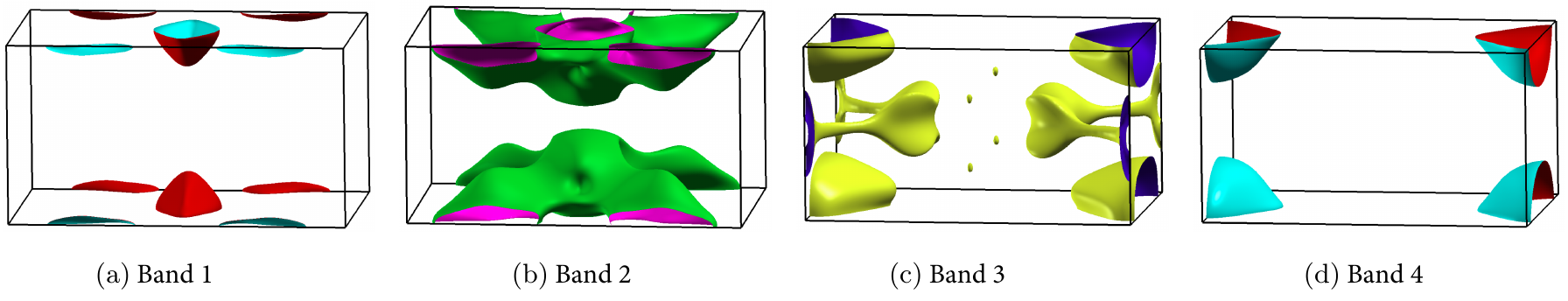}
\caption{\textbf{Fermi surfaces of NbOsSi without SOC.} Panels (a)--(d) are from a side view of the four Fermi surface sheets of NbOsSi without SOC.}
\mylabel{fig:FS}
\end{figure*}

\begin{figure}[htb]
\centering
\includegraphics[width=0.8\columnwidth]{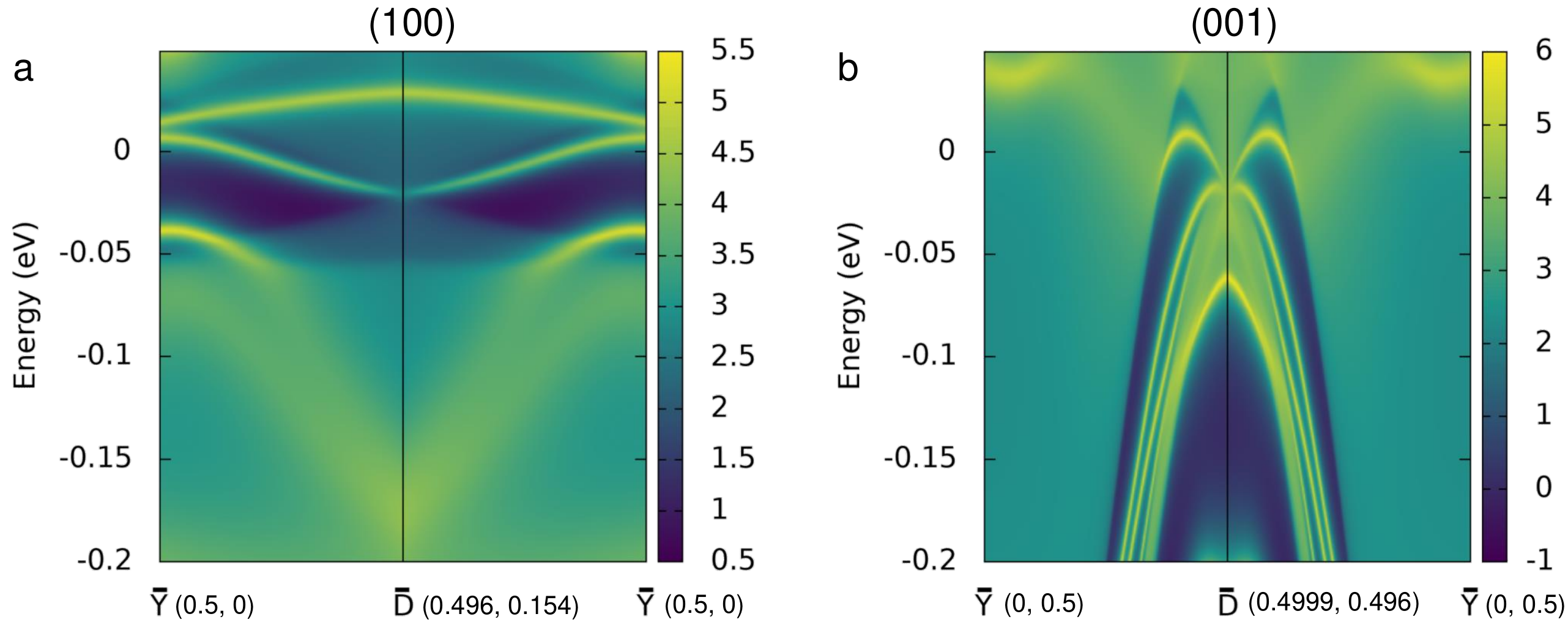}
\caption{\textbf{Surface states.} Surface states on (100) and (001) planes for NbOsSi. }
\label{fig:SurfaceStates}
\end{figure}

The coordinates of two of the Dirac points are listed in \tab{tab:DiracPoints} for both TaOsSi and NbOsSi. The other two Dirac points are just time-reversal copies of these ones. 
The dispersions close to the Dirac points of NbOsSi are shown in \fig{fig:DiracPoints}. Surface states on (100) and (001) planes for NbOsSi are shown in \fig{fig:SurfaceStates}\textbf{a} and \textbf{b}, where $\bar{D}$ is the projection of Dirac point at (0.50, 0.50, 0.16) on each plane. 

\begin{table}[h]
\caption{Positions of Dirac points for TaOsSi and NbOsSi.}
\begin{center}
\begin{tabular}{l l l l l l l l l l l l l}
 \hline
Material && \hspace{1cm} $k_{x} (2\pi/a) $ && \hspace{1cm} $k_{y} (2\pi/b) $ && \hspace{1cm} $k_{z} (2\pi/c) $  \\
\hline
TaOsSi && \hspace{1cm} 0.50 && \hspace{1cm} 0.50 && \hspace{1cm} 0.41    \\
       && \hspace{1cm} 0.50 && \hspace{1cm} 0.48 && \hspace{1cm} 0.30    \\
\hline       
NbOsSi && \hspace{1cm} 0.50 && \hspace{1cm} 0.50 && \hspace{1cm} 0.16   \\
       && \hspace{1cm} 0.50 && \hspace{1cm} 0.48 && \hspace{1cm} 0.10    \\
\hline \hline
\end{tabular}
\end{center}
\label{tab:DiracPoints}
\end{table}


\begin{figure*}[!hbt]
\centering
\includegraphics[width=0.85\textwidth]{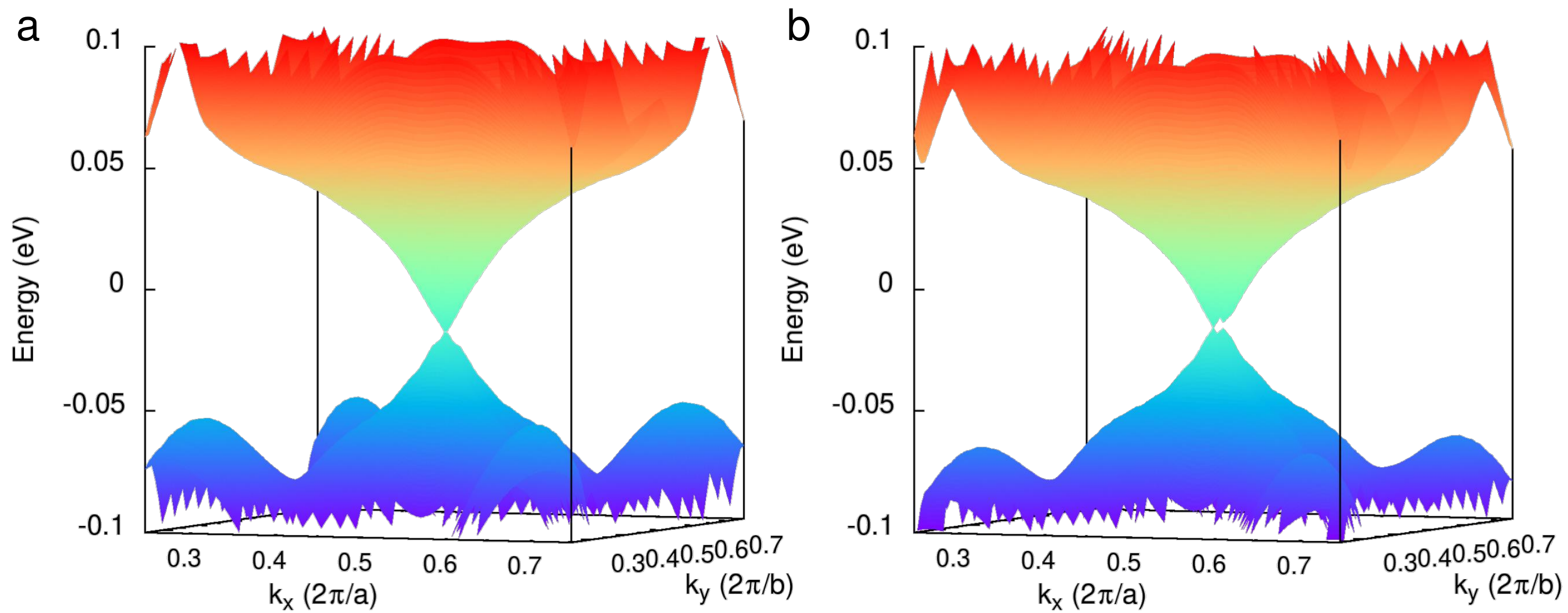}
\caption{\textbf{Dirac points of NbOsSi.} The energy dispersions near the Dirac points given in \tab{tab:DiracPoints} for NbOsSi with SOC are shown in the panels \textbf{a} and \textbf{b}.}
\mylabel{fig:DiracPoints}
\end{figure*}

\begin{figure}[htb]
\centering
\includegraphics[width=0.95\columnwidth]{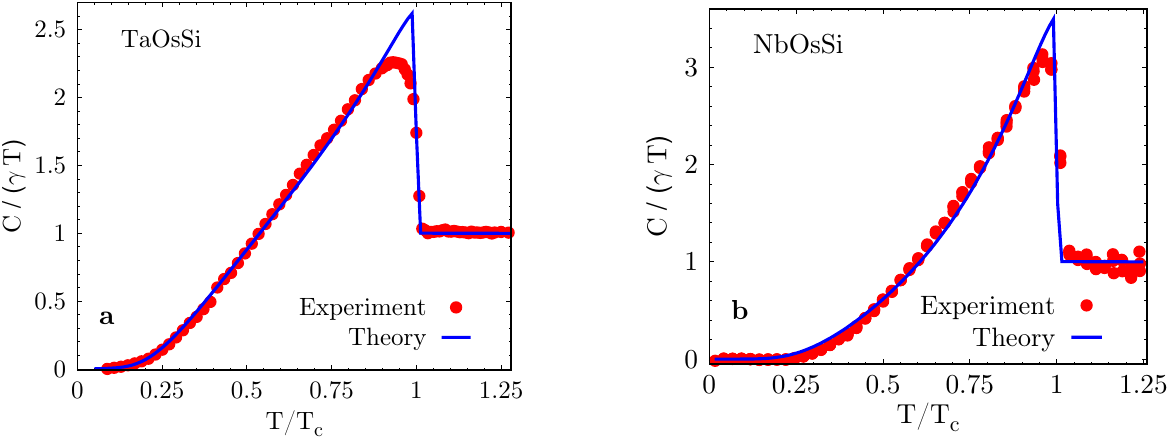}
\caption{\textbf{Fitting the specific heat in the dominant interorbital SOC limit.} Fitting the experimental electronic specific heat data considering both the intraorbital and the interorbital spin-orbit coupling with $s/t = 0.1$, $\mu/t = -3.0$ and $\pmb{\eta} = \frac{1}{\sqrt{3}}(1, e^{i\pi/100}, e^{i 101\pi/100})$. \textbf{a}) TaOsSi case with parameters $\Delta_0/(k_{\rm B} T_{\rm c}) = 2.40$, $\alpha/t = 0.15$ and $\alpha_1/\alpha = 0.001$; and \textbf{b}) NbOsSi case with parameters $\Delta_0/(k_{\rm B} T_{\rm c}) = 3.4$, $\alpha/t = 0.20$  and $\alpha_1/\alpha = 0.001$.}
\label{fig:Cv_intra_SM}
\end{figure}

\section{Properties of the INT state for a two-band toy model}
In this section, we describe the computation of the specific heat for a two-band toy model in the INT state~\cite{Ghosh2020b, Weng2016} using the Bogoliubov de-Gennes (BdG) formalism~\cite{Ghosh2020a}. In the toy model, the two rigidly shifted energy bands, which can arise from two nearly degenerate effective orbitals, have the dispersions
\beq
\epsilon_{\pm}(\bk) = \epsilon(\bk) \pm s
\eeq
where $\epsilon(\bk) = -2t [\cos(k_x) + \cos(k_y) + \cos(k_z)]$ is a generic dispersion with $t$ being a hopping parameter and $s$ being the rigid energy shift, $s/t \ll 1$.

To take into account the effect of non-zero spin-orbit coupling (SOC), found in both the materials, in a simple way, we phenomenologically consider a Rashba-type SOC, which can arise due to the local inversion symmetry breaking~\cite{Wu2017, Zhang2014, Araki2019} found in these materials. 
In general, we have both inter-orbital SOC of strength $\alpha$ and intra-orbital SOC of strength $\alpha_1$. Then the normal state Hamiltonian operator is
\beq
\hat{\cal{H}}_0 = \sum_{\bk} \hat{c}^{\dagger}_{\bk} \cdot H_0(\bk) \cdot \hat{c}_{\bk},
\eeq
defining
\beq
\hat{c}_{\bk} = \begin{bmatrix} \tilde{c}_{\uparrow, \bk} \\ \tilde{c}_{\downarrow, \bk} \end{bmatrix} \,\,\,\,\,\,{\rm with} \,\,\,\,\,\, \tilde{c}_{p,\bk} = \begin{bmatrix} c_{+, p , \bk} \\ c_{-, p, \bk} \end{bmatrix}.
\eeq
$c_{+, p , \bk}$ is an electron annihilation operator in the $\pm$ band with spin $p = \uparrow$ and $\downarrow$. The normal state Hamiltonian matrix is given by
\beq
H_0 (\bk) = \sigma_0 \otimes \begin{bmatrix} \xi_+(\bk) & 0 \\\\ 0 & \xi_{-}(\bk)\end{bmatrix} + (k_y \sigma_x - k_x \sigma_y) \otimes (\alpha \tau_x + \alpha_1 \tau_0)
\eeq
where $\xi_{\pm}(\bk) = \epsilon_\pm(\bk) - \mu$ with $\mu$ being the chemical potential and, $\sigma_0$ and $\tau_0$ are the identity matrices in spin-space and orbital-space respectively.

The interaction part of the Hamiltonian in the INT channel~\cite{Ghosh2020b, Weng2016} is
\beq
\hat{\cal{H}}_I = \frac{1}{2} \sum_{\bk} \left[\hat{c}^{\dagger}_{\bk} \cdot \hat{\Delta} \cdot {\hat{c}^{\dagger \,{\rm T}}_{-\bk}} + {\rm h. c.}\right]
\eeq
with the pairing potential in the INT state given by
\beq
\hat{\Delta} = {\Delta}_a (\pmb{\eta}\cdot \vec{\sigma})i\sigma_y \otimes i\, \tau_y,
\eeq
where $\Delta_a$ is the pairing amplitude. Then, the BdG Hamiltonian operator is
\bea
\hat{\cal{H}}_{\rm BdG} &=& \hat{\cal{H}}_0 + \hat{\cal{H}}_I \non\\
&=& \frac{1}{2} \sum_{\bk} \Psi^{\dagger}_{\bk} H_{\rm BdG}(\bk) \Psi_{\bk} + \rm{constant}
\eea
with the Nambu operators defined as
\beq
\Psi_{\bk} = \begin{bmatrix} \hat{c}_{\bk} \\ \\ { \hat{c}^{\dagger \, {\rm T}}_{-\bk}}\end{bmatrix}
\eeq
and the BdG Hamiltonian matrix is given by
\beq
H_{BdG}(\bk) = \begin{bmatrix} H_0(\bk) & \hat{\Delta} \\\\ \hat{\Delta}^\dagger & - {H_0(-\bk)}^T \end{bmatrix}.
\eeq
$H_{BdG}(\bk)$ is then diagonalized to obtain the Bogoliubov quasiparticle energy bands $E_n(\bk)$; $n = 1 \ldots 4$ which are used to compute the specific heat using the formula~\cite{Tinkham1996}
\beq
C =\sum_{n,\bk} \frac{1}{2} k_B \beta^2 \left\{E_n(\bk) + \beta \frac{\partial E_n(\bk)}{\partial \beta}\right\} E_n(\bk) \,\,{\rm sech}^2\left[\beta E_n(\bk)/2\right]
\eeq
where $\beta = \frac{1}{k_B T}$ and $k_B$ is the Boltzmann constant. The temperature dependence comes in the quasiparticle energy bands considering that the temperature dependence comes only from the pairing amplitude~\cite{Carrington2003} as
\beq\label{eqn:temp_dep}
\Delta_a(T) = \Delta_0 \tanh\left[1.82\left\{1.018\left(T_{\rm c}/T-1\right)\right\}^{0.51}\right]
\eeq
with $\Delta_a(0) \equiv \Delta_0$. We ignore any weak temperature dependence in the $\bq$-vector.

To fit the specific heat data, we first fix the $T_{\rm c}$ of a particular material from experiments and then treat $\Delta_0/(k_B T_c)$, $\pmb{\eta}$, $\alpha/t$ and $\alpha_1/\alpha$ as fitting parameters. So, in general we have four fitting parameters. However, we can only fit the experimental specific heat data well in the limit $\alpha_1/\alpha \ll 1$, i. e. dominant interorbital SOC, without changing the topology of the resulting Fermi surfaces qualitatively. As seen from the example shown in \fig{fig:Cv_intra_SM} in this limit we can fit the specific heat data for both the materials quite well in the weak coupling limit and a nonzero $\alpha_1$ does not change the physics qualitatively as long as it is much smaller than $\alpha$. Hence, we only consider the case $\alpha_1 = 0$ in the main text for simplicity.

\clearpage

\setcounter{page}{1}
\setcounter{figure}{0}


\end{document}